\documentstyle[12pt,prd,aps]{revtex}
\begin{document}
\title{Recycling Universe} 
\author{Jaume Garriga} \address{IFAE, Departament di Fisica,
Universitat Autonoma de Barcelona,\protect\\ 08193 Bellaterra
(Barcelona), Spain} 
\author{Alexander Vilenkin} \address{Institute of Cosmology, Department of
Physics and Astronomy,\protect\\ Tufts University, Medford, MA
02155, USA} 
\maketitle
\begin{abstract}

If the effective cosmological constant is non-zero, our observable 
universe may enter a stage of exponential 
expansion. In such case, regions of it may tunnel back to the false vacuum of 
an inflaton scalar field, and inflation with a high expansion rate may 
resume in those regions. An ``ideal'' eternal 
observer would then witness an infinite succession of cycles
from false vacuum to true, and back. Within each cycle, the 
entire history of a hot universe would be replayed. If there were several 
minima of the inflaton potential, our ideal observer would visit
each one of these minima with a frequency which depends on the shape
of the potential. We generalize the formalism of stochastic inflation
to analyze the global structure of the universe when 
this `recycling' process is taken into account. 

\end{abstract}
\pacs{1}

\section{Introduction}

Inflationary models are designed to produce a universe which is
sufficiently homogeneous on all observable scales \cite{Review}.
However, on much larger scales the universe is expected to be
extremely inhomogeneous.  The evolution of the field $\phi$, whose
vacuum energy drives inflation, is influenced by quantum fluctuations.
These fluctuations can be pictured as a random walk of $\phi$
superimposed on its slow roll down the slope of its potential.  
As a result, thermalization of the vacuum energy does not occur
simultaneously everywhere in the universe, and at any time there are
parts of the universe that are still inflating \cite{AV83,Linde86}.

On very large scales, the universe is expected to consist of isolated
thermalized regions embedded in the inflating background.  The
boundaries of the thermalized regions expand into this background, and
new regions are constantly being formed, but the high expansion rate
of the intervening inflating domains prevents these regions from
filling up the universe.  Thermalization inevitably occurs at any
given co-moving location, and the co-moving volume of the inflating
regions decreases exponentially with time.  At the same time, the
physical volume of these regions is exponentially growing.  The
geometry of the inflating regions is that of a self-similar fractal of
dimension $d<3$ \cite{Aryal}.  It is illustrated in Fig.1 for the case
of ``open''inflation, where the false vacuum decay occurs through
bubble nucleation \cite{Open,LM95}.  For ``new'' or ``chaotic''
inflation the picture would be similar, except the thermalized regions
would have irregular shapes.

In the present paper we are going to argue that this picture of the
superlarge-scale structure of the universe can be significantly
modified by quantum fluctuations that bring localized parts of already
thermalized regions, such as our observable universe,
back to the inflating false-vacuum state.  The
modification is particularly important in models where the
post-thermalization true vacuum is characterized by a positive vacuum
energy (cosmological constant).  In this case the thermalized regions
asymptotically approach de Sitter geometry, and the rate of
fluctuations back to the false vacuum (per unit spacetime volume)
approaches a constant.  Even with an exceedingly small rate, the
probability for true vacuum to survive at any co-moving location is
exponentially decreasing with time.  Hence, (almost) all the co-moving
volume of thermalized regions will eventually be recycled back to the
inflationary phase.  Each nucleated false vacuum region will serve as
a seed for a new eternally inflating domain, whose internal structure
will resemble that shown in Fig.1.  The thermalized regions formed in
this domain will in turn produce new false vacuum seeds, {\it etc}.  
We call this kind of model a recycling universe.

Quantum nucleation of regions with a higher energy density cannot
occur from a flat-spacetime vacuum characterized by a vanishing
cosmological constant: such processes are forbidden by energy
conservation.  However, upward fluctuations of this kind can occur in
an expanding cosmological background, and have been previously
discussed 
by a number of authors \cite{HM,AV83,Linde86,Starob86,GL86,Linde91,LW}.  
The most relevant
for our purposes here is the paper by Lee and Weinberg \cite{LW} who
considered a model of a scalar field $\phi$ with a potential $V(\phi)$
shown in Fig.2.  Note that both false and true vacua have positive
energy densities, $\rho_f>\rho_t>0$.  
It has been known for some time that the high-energy
false vacuum at $\phi=0$ can decay by nucleation of true vacuum
bubbles.  The corresponding instanton (``bounce'') 
has been found by Coleman and De Luccia \cite{CdL}.  The bubble
nucleation rate is given by
\begin{equation}
\Gamma_{f\to t}= A\exp [-S_b +S_f],
\label{bouncet}
\end{equation}
where $S_b$ is the bounce action and $S_f=-3/8\rho_f$ is the action of
the Euclideanized false-vacuum de Sitter space (we use Planck units
throughout the paper).
Lee and Weinberg conjectured that the same
instanton also describes the inverse process of true vacuum decay,
where false vacuum bubbles nucleate in a true vacuum background.  The
nucleation rate suggested by Eq.(\ref{bouncet}) is
\begin{equation}
\Gamma_{t\to f}= A\exp [-S_b +S_t]
\label{bouncef}
\end{equation}
with $S_t=-3/8\rho_t$.  Lee and Weinberg argued that the
pre-exponential factors in Eqs. (\ref{bouncet}) and (\ref{bouncef})
are the same.
These conjectures were later verified \cite{Garriga} in the case of
$(1+1)$-dimensional universes, where bubble nucleation can be
identified with the production of particle-antiparticle pairs.
Note that the rate (\ref{bouncef}) vanishes if the true vacuum has
zero energy.

As it stands, the potential in Fig.2 is not suitable for inflationary
cosmology.  This potential has no slow-roll region, so most of the
vacuum energy remains in domain walls and never gets thermalized.  
We shall see, however, that models incorporating both realistic
inflation and true vacuum decay can be constructed by a trivial
modification of ``open'' inflationary models.
Moreover, we shall argue that nucleation of inflating regions is
possible even with the simplest slow-roll potentials, for which the
Coleman-de Luccia instanton does not exist.

The recycling nature of inflationary universe may have important
implications for the question of whether or not the universe had a
beginning in time.  As we already mentioned, inflation is generically
eternal to the future, so it is natural to ask if the inflationary
models can be continued into the infinite past, resulting in a
``steady-state'' non-singular cosmology.  This possibility was
discussed in the early 80's, soon after the inflationary scenario was
proposed, with the conclusion that the idea could not be implemented
in the simplest model in which
the inflating universe is described by an exact de Sitter space
\cite{Linde83,AV83}.  A more general proof of impossibility of steady state
inflation was given in Refs. \cite{BV94,Borde94,BVrev}, but we shall see later in this
paper that some of the assumptions made in the proof do not apply in
the case of a recycling universe.  The question of the necessity of
the beginning is therefore re-opened.

Recycling may also be relevant to the question of making predictions
in an inflationary universe. Recently, there have been a number of 
attempts \cite{GBL95,LLM94,AV94,AV95,VW97,MSW97}
to find probability distributions for cosmological parameters such as
the effective cosmological constant $\Lambda$ or the density parameter 
$\Omega$. These 
``predictions'' are based on the principle of mediocrity \cite{AV94,Foot5},
by which we are most likely to live in the most abundant 
type of civilization that can result from the thermalization of a fase vacuum.
However,  in the inflationary universe, there will be an infinite number
of infinite thermalized regions, and one faces 
the difficulty of comparing infinities \cite{LLM94,GBL95,WV96}. 
Regularization procedures
were introduced in \cite{AV95,LM96} to deal with this problem, but these 
cannot be directly applied to a recycling universe. 

The paper is organized as follows.  In the next Section we shall give
some examples of inflationary models which allow nucleation of false
vacuum bubbles.  The geometry of the nucleated bubbles will be
analyzed in Section 3. In
Section 4 we
shall discuss the implications of the recycling universe model for the
question of the beginning of the universe.  The
superlarge-scale structure of a recycling universe will be studied in
Section 5 using the methods of stochastic inflation.  The issue of
predictions will be discussed in Section 6, and our conclusions
will be summarized in Section 7.

\section{Models}

In all realistic inflationary models, the potential of the inflaton
field $\phi$ is required to have a sufficiently flat 
slow-roll region in which
\begin{equation}
|V''(\phi)|\ll H^2.
\label{slowroll}
\end{equation}
Here, $H$ is the expansion rate and $H^{-1}$ is the horizon size
corresponding to the vacuum energy $V(\phi)$,
\begin{equation}
H^2=8\pi V(\phi)/3.
\label{H}
\end{equation}
On the other hand, Coleman-de Luccia-type solutions for vacuum bubbles
exist only when the potential is sufficiently curved near the barrier
separating true and false vacua \cite{Rama}, 
\begin{equation}
|V''(\phi)|\gtrsim H^2.  
\label{cond}
\end{equation}
The meaning of this condition is easy to
understand.  The bubble wall thickness is $\delta\sim
|V''|^{-1/2}$, and if (\ref{cond}) is not satisfied, then the wall is
much thicker than the horizon.  Such walls cannot exist as coherent
structures and are spread by the expansion of the universe.  False
vacuum  bubbles of Coleman-de Luccia type are, therefore, impossible 
if the slow roll condition 
(\ref{slowroll}) is valid everywhere in the inflationary range of
$\phi$. 

A similar problem arises in the ``open'' inflation scenario, where
false vacuum decay through bubble nucleation is followed by a period
of slow roll in bubble interiors \cite{Open,LM95}.  
One way to deal with this
problem is to consider a two-field model, with one field doing the
tunneling and the other doing the slow roll \cite{LM95}. 
The potential
can be chosen as 
\begin{equation}
U(\chi,\phi)=V_1(\chi)+\chi^2 V_2(\phi).
\label{LM}
\end{equation}
Here, $\chi$ is the tunneling field and the potential $V_1(\chi)$ has
the form as in Fig.2 with a metastable minimum at $\chi=0$ and a true
minimum at $\chi=\eta_\chi$.  The full
potential $U(\chi,\phi)$ is independent of $\phi$ (has a flat
direction) at $\chi=0$, and
as a result the expansion rate  in the false vacuum is also
independent of $\phi$.  The potential $V_2(\phi)$ is assumed to have
a slow-roll range and a minimum at $\phi=\eta_\phi$ with
$V_2(\eta_\phi)=0$, at which thermalization eventually occurs.
In a variant of this model \cite{LM95}, the two fields
can be taken to represent the radial and angular parts of a single
complex field, $\Phi=\chi e^{i\phi}$.

The stage for open inflation is set by inflating false vacuum with
energy density $\rho_f=V_1(0)$.  Nucleating bubbles expand into this 
background, but because of the high expansion rate of the intervening
false vacuum regions, bubble collisions are rare.  The interior
geometry of each bubble is that of an open Robertson-Walker universe.
The bubbles have different initial 
values of $\phi$, and if this value falls in the slow-roll
range of the potential, then there is a period of inflation inside the
corresponding bubble.  Inflation is followed by thermalization and
standard cosmological evolution, but since we assumed a non-zero
cosmological constant, the bubble interiors are eventually dominated
by the true vacuum energy, $\rho_t=V_1(\eta_\chi)$.  False vacuum bubbles
will now be formed in the true vacuum background, resulting in the
endless succession of stages of the recycling universe.

It should be noted that tunneling back to the false vacuum can occur
not only from the true vacuum $(\chi,\phi)=(\eta_\chi,\eta_\phi)$, but
also from the slow-roll, as well as radiation and matter-dominated
periods.  In fact, the rate of false vacuum bubble nucleation is
expected to be the highest during the slow-roll inflation [due to the
higher energy density at that time, see Eq.(\ref{bouncef})].  
However, since all these periods last only for 
a finite time, and the rate of false vacuum bubble formation
is extremely low, only a tiny fraction of the co-moving volume will be
affected by such processes.  On the other hand, the
true-vacuum-dominated stage persists indefinitely, and practically all
the co-moving volume will be recycled by bubbles nucleating in the
true vacuum.

An alternative to the two-field model (\ref{LM}) is a model of a
single scalar field $\phi$ with a potential of the form shown in Fig.3
\cite{Open}.  The false vacuum at $\phi=0$ is separated from the
slow-roll region by a sharp barrier.  (The coexistence of flat and
highly curved regions in the same potential is a somewhat unnatural
feature of this model).  The field $\phi$ tunnels through the barrier
and after a period of slow roll, ends up in the true vacuum, which we
assume to have a non-zero energy density, $\rho_t>0$.  An important
difference of this model from that of Eq.(\ref{LM}) is that now, in
order to tunnel from true to false vacuum, the field $\phi$ has to go
across the whole slow-roll region.  False vacuum bubbles will
therefore consist of a false vacuum core surrounded by a domain wall,
which is in turn surrounded by layers of slow roll, radiation, and
matter-dominated regions (see Fig.4).  

Intuitively, we would expect that the nucleation of such a complicated 
structure
should be extremely unlikely, and thus the nucleation rate 
of false vacuum bubbles in this one-field model should be strongly
suppressed compared to the two-field model (\ref{LM}).  However,
according to Eq.(\ref{bouncef}) this would not be so.  The
bounce solution depends only on the form of the potential $V(\phi)$ in
the vicinity of the barrier. Thus the tunneling rate given by the
expression
(\ref{bouncef}) is essentially independent of what happens in the
slow-roll region and would not be much different if that region was
absent.  It should be remembered though 
that the use of Euclidean methods in
de Sitter space has never been justified from first principles, and
therefore the results obtained using these methods should be taken
with caution \cite{Starob86,GL86,Linde91}.  This issue needs further 
investigation, but we will not attempt to address it in the present paper.  
The specific value of false vacuum bubble nucleation rate will not be
important for our conclusions, as long as this rate is non-zero.

If the barrier in the inflaton potential is too wide to satisfy the
condition (\ref{cond}), then the Coleman-de Luccia instanton does not
exist.  However, there is always a homogeneous Hawking-Moss instanton
\cite{HM} in which $\phi$ takes the value $\phi_b$ 
corresponding to the top of the barrier.  This instanton
is usually interpreted as describing quantum tunneling from false
vacuum to the top of the barrier in a horizon-size region.
(Coleman-de Luccia instanton reduces to that of Hawking and Moss as
the barrier width is increased).  In the spirit of Lee and Weinberg
\cite{LW}, 
we can interpret the same instanton as describing tunneling from 
true vacuum to the top of the barrier.  The corresponding nucleation
rate is 
\begin{equation}
\Gamma\propto \exp [-S_{HM}+S_t],
\label{bouncehm}
\end{equation}
where $S_{HM}=-3/8\rho_b$ and $\rho_b=V(\phi_b)$.

In models of ``new'' inflation, the generic potential is illustrated
in Fig.5.  There is no barrier in this case, but still there is a
Hawking-Moss instanton with $\phi_b$ corresponding to the maximum of
the potential.  There are also approximate homogeneous instanton 
solutions with  $\phi$ sufficiently close to the maximum.  Such
approximate instantons also exist in models of ``chaotic'' inflation
where the potential may have no maxima.  A constant
field $\phi$ is a good approximation as long as the evolution of
$\phi$ is slow on the Hubble scale $H^{-1}$, that is, in the slow roll
range.  The Euclidean expression for the tunneling rate from true
vacuum to $\phi$ in this range is given by Eq.(\ref{bouncehm}) with
$S_{HM}=-3/8V(\phi)$.  

Once again, we find the formula (\ref{bouncehm}) somewhat
suspicious, and emphasize the need for a derivation of the nucleation
rate without relying on Euclidean methods.
Linde \cite{Linde91} has given an estimate of the probability of 
tunneling to false vacuum on rather general grounds. In de Sitter 
space, a field $\phi$ fluctuates on scales bigger or comparable to the Hubble 
radius 
around a local minimum $\phi_t$ of the potential with amplitude given by
\cite{BD}
$$
\sigma^2=<(\phi-\phi_t)^2>\approx {3H^4 \over 8\pi^2 m^2}.
$$
Strictly speaking, the result
is only true for a free field and for small fluctuations, so that
the mass $m$ and the expansion rate $H$ are well defined. Extrapolating 
to the case of an interacting field and ignoring the gravitational 
backreaction of the fluctuations on the expansion rate, the 
probability that a region of size $\sim H^{-1}$ will tunnel from
$\phi_t$ to a different value $\phi$ sufficiently close to the maximum 
of $V(\phi)$ can be estimated as
$$
P\propto \exp[-(\phi-\phi_t)^2/2 \sigma^2].
$$
For the case of a quadratic potential with $m^2<< H^2$ and 
$V(\phi)-V(\phi_t)\ll V(\phi_t)$,
the exponent in the previous expression reproduces 
the exponent $[-S_{HM}+S_t]$ which appears in Eqn. (\ref{bouncehm}),
with $S_{HM}=-3/8V(\phi)$ \cite{Linde91}.

In any case, we expect
the rate to be non-zero in the general case.   The reason is simply that
the nucleation is not forbidden by any conservation laws, and thus
should have a non-zero probability. 

\section{False vacuum bubbles}

To study the geometry of false vacuum bubbles, we shall
first assume that the bubble wall thickness is much
smaller than all other relevant dimensions of the problem.  
The wall can then be treated as infinitely thin, and the spacetime
regions on the two sides of the wall are de Sitter spaces of different
vacuum energy.  We now briefly review some properties of de Sitter
space. 

It is well known that de Sitter space can be pictured as a hyperboloid
embedded in a flat 5-dimensional spacetime,
\begin{equation}
{\bf \zeta}^2 +w^2-\tau^2=H^{-2}.
\label{hyperboloid}
\end{equation}
where ${\bf\zeta}=(\zeta^1,\zeta^2,\zeta^3)$ is a 3-vector.
A section of the hyperboloid by the $w\tau$-plane is shown in Fig.6.
The Euclideanized de Sitter space, which is used for constructing
instantons, is obtained by analytic continuation $\tau=i\tau_E$,
\begin{equation}
{\bf \zeta}^2 +w^2+\tau_E^2=H^{-2}.
\label{sphere}
\end{equation}
Geometrically, this is a 4-sphere of radius $H^{-1}$.

It will be convenient to use the Robertson-Walker flat coordinates in
which the de Sitter metric takes its most familiar form,
\begin{equation}
ds^2=dt^2-e^{2Ht}d{\bf x}^2,
\label{desitter}
\end{equation}
These coordinates are related to the hyperboloid coordinates by 
\begin{equation}
\tau=H^{-1}\sinh (Ht)+{1\over{2}}H{\bf x}^2e^{Ht},
\end{equation}
\begin{equation}
w=H^{-1}\cosh (Ht)-{1\over{2}}H{\bf x}^2e^{Ht},   
\label{hyp-ds}
\end{equation}
\begin{equation}
{\bf \zeta}={\bf x}e^{Ht},
\end{equation}
which can be inverted to give
\begin{equation}
t=H^{-1}\ln [H(w+\tau)],~~~~{\bf x}={H^{-1}{\bf \zeta}\over{w+\tau}}.
\label{ds-hyp}
\end{equation}
Constant-$t$ surfaces are obtained as intersections of the hyperboloid
with null hyperplanes $w+\tau=const$ (see Fig.6).  The surface
$t=-\infty$ corresponds to $w+\tau=0$, and thus the coordinate system
(\ref{desitter}) covers only half of de Sitter space.

Let $H_f^{-1}$ and $H_t^{-1}$ be de Sitter horizons corresponding
respectively to the false and true vacuum
energy densities, $\rho_f$ and $\rho_t$.  Clearly,
$\rho_t< \rho_f$ and $H_f^{-1}< H_t^{-1}$.  The thin wall 
approximation requires that the wall thickness be much smaller than
$H_f^{-1}$.  In this case, the Coleman-de Luccia instanton for bubble
nucleation can be obtained by matching two 4-spheres of radii
$H_f^{-1}$ and $H_t^{-1}$ (see Fig.7).  The two spheres are joint at a
3-sphere which represents the Euclideanized worldsheet of the domain
wall.  Its radius $R_0$ is determined by $\rho_f$, $\rho_t$, and the wall
tension $\sigma$ \cite{BKT}.  The 5-dimensional coordinates can always
be chosen so that this worldsheet lies in a plane of constant $w$.
In the figure it is $w=-D_t$, where 
\begin{equation}
D_t= (H_t^{-2}-R_0^2)^{1/2}.
\label{dt}
\end{equation}
The Lorentzian evolution of the bubble is given by two hyperboloids
similarly matched along a constant-$w$ plane (Fig.8).
Descriptions of both true and false vacuum bubbles can be obtained
with an appropriate slicing of this spacetime by equal-time surfaces.

In the case of false vacuum bubbles, equal-time surfaces can be chosen
to be the surfaces $w+\tau= const$ (Fig.8).  Then, each constant-$t$
slice consists of a spherical region of false vacuum embedded in an
infinite, spatially-flat region of true vacuum. Since the
spatial geometry of these slices is flat both inside and outside the
bubble, the volume that is removed from true vacuum by the appearance of
the bubble is equal to the volume of false vacuum which replaces it
(this would not be the case if we used closed spatial sections, for instance). 
We can use the
coordinates (\ref{desitter}) with $H=H_t$ to describe the exterior
true-vacuum region.  The wall worldsheet is at $w=-D_t$, and from
Eq.(\ref{hyp-ds}) the radius of the bubble at time $t$ is
\begin{equation}
R_f^2(t)=H_t^{-2}(e^{2H_tt}+2D_tH_te^{H_tt}+1),
\label{rf}
\end{equation}
where $R(t)=|{\bf x}(t)|e^{H_tt}$.  We see that the radius approaches
the horizon size $H_t^{-1}$ at $t\to -\infty$. 
The bubble wall accelerates in the direction of the false
vacuum, so that its co-moving radius $|{\bf x}(t)|$ is contracting,
but the physical radius grows exponentially due to the expansion of
the universe.  A conformal diagram for the bubble spacetime is shown
in Fig.9.

Since $R(t)$ is a monotonically growing function of time, there is no
``bounce'' moment at which one can say that the nucleation occurs.
The situation here is similar to that for nucleation of topological
defects in de Sitter space \cite{Rama}, and as in the latter case, we
shall regard the bubble (\ref{rf}) as ``formed'' at $t\sim 0$, when
its radius begins to grow exponentially.  

For the trajectory given in Eqn. (\ref{rf}), the region of true vacuum
which has been removed and replaced by true vacuum was centered at
the point $(t=0,{\bf x}=0)$ (see Figs. 6 and 8). We shall refer to this point
as the center of symmetry of the bubble trajectory. (Notice that this point may
not belong to the actual classical spacetime, because it is precisely in
the region where bubble nucleation takes place,
but it does belong to the five-dimensional embedding space.) 
Performing Lorentz  transformations in the embedding space, 
we can obtain bubbles whose center of symmetry is at any time $t=t_0$ 
and at any location. 
This gives \begin{equation}
R_f^2(t)=H_t^{-2}[e^{2H_t(t-t_0)}+2D_tH_te^{H_t(t-t_0)}+1].
\label{rft0}
\end{equation}  
We can think of these as bubbles ``formed'' at $t\sim t_0$.
The asymptotic behavior of the bubble radius at large times is
\begin{equation}
R_f(t)\approx H_t^{-1}e^{H_t(t-t_0)}, ~~~~~~~~~~ t-t_0\gg H_t^{-1}.
\label{rfas}
\end{equation}  

To describe a true vacuum bubble in a false vacuum background, we
choose equal-time surfaces to be $w-\tau=const$.  Using the
coordinates (\ref{desitter}) with $H=H_f$ to describe the exterior 
of the bubble, we find that the bubble radius at time $t$ is given by
\begin{equation}
R_t^2(t)=H_f^{-2}(e^{2H_ft}-2D_fH_fe^{H_ft}+1),
\label{rt}
\end{equation}
where
\begin{equation}
D_f= (H_f^{-2}-R_0^2)^{1/2}.
\label{df}
\end{equation}
In contrast to the false-vacuum bubble case, the radius (\ref{rt}) has
a minimum, $R_{min}=R_0$, at $t_n =H_f^{-1}\ln(D_fH_f)$,
and we can regard this time as the moment of bubble nucleation.
Eq. (\ref{rt}) can then be rewritten as
\begin{equation}
R_t^2(t)=D_f^2[e^{2H_f(t-t_n)}-2e^{H_f(t-t_n)}]+H_f^{-2},
\label{rttn}
\end{equation}
with the late-time behavior
\begin{equation}
R_t(t)\approx D_fe^{H_f(t-t_n)}, ~~~~~~~~~~ t-t_n\gg H_f^{-1}.
\label{rtas}
\end{equation}
Note, however that
\begin{equation}
R_t(t)\approx H_f^{-1}e^{H_f(t-t_0)}, ~~~~~~~~~~ t-t_0\gg H_f^{-1},
\label{rtas2}
\end{equation}
where we denote by $t_0$ the 
location of the center of symmetry of the bubble wall as seen from the
outside [see the discussion around equation (\ref{rft0})].

Our main interest in this paper is in models with
$\rho_t\ll\rho_f$ .  In such models, the
radius of false vacuum bubbles is $R(t)>H_t^{-1}\gg H_f^{-1}$, 
and thus the
thin wall approximation can be used to describe the bubble evolution
even when the wall thickness is $\delta\sim H_f^{-1}$.  
(Note however that in this case the thin wall approximation breaks
down for the instanton itself and for the early evolution
of true vacuum bubbles).  

In a single-field model of open inflation, with a potential as in Fig. 3, 
the conformal diagram for a false vacuum bubble is shown in Fig.10.  

So far in this Section we assumed that the true vacuum has a positive
energy density.  If the vacuum energy is in fact zero, then the
horizon radius in thermalized regions keeps growing with time, and
false vacuum bubbles eventually come within the horizon.  The bubbles
are then seen as black holes from the outside.   
The bubble nucleation in this case is similar to the
quantum creation of baby universes, as discussed in
Refs. \cite{FGG,FMP,Linde91}.  Black holes eventually evaporate and
baby universes pinch off.

\section{Did the universe have a beginning?}

Assuming that some rather general conditions are met, it was shown in
Ref. \cite{BV94} that inflationary models cannot be geodesically complete
to the past, that is, they require some sort of a beginning.  The
assumptions that lead to this result are the following.

A. The universe is causally simple \cite{Foot1}.

B. The universe is open.

C. The null convergence condition \cite{Foot2}.

D. The finite past-volume difference condition.

The first two of these assumptions do not appear to be crucial for the
proof, and extensions of the theorem have been obtained to some closed
universes \cite{Borde94} and to some universes with a more complicated
causal structure \cite{BVrev}.  

The null convergence condition is closely related to the weak energy
condition, which requires that the energy density is non-negative when
measured by any observer.  Classically, this is satisfied by all known
forms of matter, including a relativistic scalar field, but violations
of the null
convergence condition are possible as a result of quantum
fluctuations.  Such violations tend to occur in the inflating regions
of spacetime whenever quantum fluctuations result in a local increase
of the expansion rate, $dH/dt >0$ \cite{BV97}.  They are sufficient to
invalidate the theorem in models characterized by a substantial
variation of $V(\phi)$ in the range of $\phi$ where quantum
fluctuations are non-negligible.  This includes all models of
``chaotic'' inflation, but not some open and ``new'' inflationary
models. 

Turning now to the effects of recycling, we shall argue that they
invalidate both assumptions C and D, so that the theorem as it stands
does not apply to any inflationary model.  In the case of assumption
C, the reason is the same as before: quantum fluctuations from
thermalized regions back to the inflating phase increase the expansion
rate in the affected regions of space, and the null convergence
condition is violated.

The finite past-volume difference condition ($D$) can be formulated as
follows.  Given a
point $P$ in the inflating region and a point $Q$ to the past of $P$, 
consider the difference of their pasts.  This is a spacetime region
including all points to the past of $P$, but not of $Q$.
The condition $D$  
requires that the spacetime volume of this region should be finite.
The original motivation for this condition was based on the picture of
eternally inflating universe without recycling, as illustrated in
Fig.1.  It can be shown that thermalization surfaces, which separate
inflating and thermalized regions of spacetime, are spacelike surfaces
\cite{BV94}.  Therefore, if $Q$ is a point in an inflating region, then,
disregarding recycling, all points in its past must also lie in the
inflating region.  For inflation to persist from $Q$ to $P$, no
thermalized regions should be formed in the difference of the pasts of
the two points.  Now, it seems plausible that there is a zero
probability for no thermalized regions to form in an infinite
spacetime volume.  Then it follows that condition $D$ is
necessary for inflation to persist in the future time direction.  
This condition is difficult to satisfy, since the spacetime region
representing the difference of the pasts of the two points has an
infinite extent along the past-directed null geodesics.

In a recycling universe, this logic does not apply, since 
points in inflating regions can have thermalized regions in
their past.  In this case, the
spacetime region relevant for persistence of inflation between a pair
of points is not the entire difference of their pasts, but 
only the part of this difference which lies in the same inflating
region as the two points.  In other words, it is the part of the
difference of the pasts which is to the future of the 
``nucleation surface'' (see Fig.9).  The volume of this region is
obviously finite.  Hence, Assumption $D$ is not suitable for a
recycling universe.

In models with a vanishing true vacuum energy, only a small fraction
of thermalized volume gets recycled.  However, there still appears to
be a possibility that the universe has a nested structure, with all
inflating regions originating as quantum fluctuations inside
thermalized regions.

We thus conclude that the theorems of Refs. \cite{BV94,Borde94,BVrev} 
no longer apply
when the recycling nature of the universe is taken into account.  This
may open the door for constructing non-singular, steady-state
inflationary models.  We emphasize, however, that our analysis does
not imply that such models do in fact exist.  It has been
argued in Refs. \cite{AV92,BV97} that inflation, when continued to the past,
is necessarily preceeded by a period of contraction, as in the exact
de Sitter space.  During this period, the thermalized regions would
merge, the density perturbations would grow very fast, and the universe
would rapidly reach a grossly inhomogeneous state from which it is not
likely to recover.  The arguments in \cite{AV92,BV97} do not rely on weak
energy or finite past-volume difference conditions, and may possibly
be extended to the case of a recycling universe.  These arguments,
although suggestive, fall short of a proof, and the problem requires
further investigation.

\section{Stochastic formalism}

A quantitative description of the recycling universe can be given
using the formalism of stochastic inflation developed in
Refs. \cite{AV83,Starob86,LLM94}.  A straightforward extension of this
formalism  will be
required, and to introduce the necessary modifications, 
we shall first consider the
Lee-Weinberg model \cite{LW} with a potential as in Fig.2.  

\subsection{Lee-Weinberg model}

Consider an ensemble of co-moving observers whose world lines are
orthogonal to some spacelike hypersurface $\Sigma$. Let $p_f(\tau)$
and $p_t(\tau)$ be the fractions of observers in false and true
vacuum, respectively, 
\begin{equation}
p_f(\tau)+p_t(\tau)=1,
\label{norm}
\end{equation}
where $\tau$ is the proper time along the
observer's world lines measured from their intersection with $\Sigma$.
The time evolution of $p_f$ and $p_t$ is described by the system of
equations
\begin{equation}
dp_f/d\tau=-\kappa_f p_f+\kappa_t p_t,
\label{eqpf}
\end{equation}
\begin{equation}
dp_t/d\tau=-\kappa_t p_t+\kappa_f p_f.
\label{eqpt}
\end{equation}
Here, $\kappa_f$ ($\kappa_t$) is
the probability, per unit time, for an observer who is presently in
the false (true) vacuum
to find himself within a true (false) vacuum bubble.

>From Eq. (\ref{rttn}), we see that a false-vacuum observer will 
be affected only by bubbles nucleating within a sphere of radius $D_f$
centered on that observer.  The bubbles take time $\sim
H_f^{-1}$ to traverse this distance, but in the stochastic inflation
formalism we shall be interested in quantities smeared over a spacetime
scale $\sim H^{-1}$, so we shall disregard this time delay and write
\begin{equation}
\kappa_f\approx\Gamma_f^{(n)}{4\pi\over{3}}D_f^3,
\label{kappaf}
\end{equation}
where $\Gamma_f^{(n)}$ is the rate of bubble nucleation per unit spacetime
volume.

In an expanding universe, however, the rate of nucleation 
per unit spacetime volume has to be defined with some care.
This is because this rate depends on  what we choose as the 
nucleation time, which is not always sharply defined (especially when the 
size of bubbles becomes comparable to the Hubble radius.)
For instance, we can change our definition of nucleation time from $t_n$ 
to $t_0$, where $t_n$ is defined by Eqn (\ref{rttn}) as the time at which the
physical radius of the bubble reaches its minimum value $R_0$, and $t_0$
is defined as the center of symmetry of the bubble wall trajectory
as seen from the outside (as explained in Sect. III).
In changing the definition of nucleation time,
we must simultaneously change the definition of nucleation rate per unit 
voulme, because the physical volume has increased by the amount 
$\exp[3H_f(t_0-t_n)]$ in the intervening time. Distinguishing by their
superindex the 
rates associated with both choices of nucleation time, we have
$$
\Gamma_f^{(n)}=\Gamma_f^{(0)} \exp[3H_f(t_0-t_n)].
$$
Hence, Eqn. (\ref{kappaf}) can be rewritten as
\begin{equation}
\kappa_f\approx\Gamma_f^{(0)}{4\pi\over{3}}H_f^{-3}.
\label{kappaf2}
\end{equation}

In the case of false vacuum bubbles the physical radius grows monotonically 
with time, so the analogue of $t_n$ does not exist. We shall adopt the 
convention identifying the nucleation time with
$t_0$ in Eq. (\ref{rft0}).  Then the co-moving region
affected by the bubble is a sphere of radius $H_t^{-1}$, and we can
write 
\begin{equation}
\kappa_t\approx\Gamma_t^{(0)}{4\pi\over{3}}H_t^{-3},
\label{kappat}
\end{equation}
where $\Gamma_t^{(0)}$ is the corresponding nucleation rate.  We note that
the radius of the affected region and the rate would both
be modified with a different choice of nucleation time, while $\kappa_t$
would remain unchanged.

The solution of Eqs.(\ref{eqpf}),(\ref{eqpt}) is
\begin{equation}
p_f(\tau)=p_f^{(0)}+Ae^{-b\tau},
\end{equation} 
\begin{equation}
p_t(\tau)=p_t^{(0)}-Ae^{-b\tau},
\end{equation} 
where the constant $A$ is determined by the initial conditions,
$b=\kappa_f+\kappa_t$, and $p_f^{(0)}$, $p_t^{(0)}$ is a stationary
distribution defined by
\begin{equation}
{p_f^{(0)}\over p_t^{(0)}}={\kappa_t\over \kappa_f}={H_f^3\over H_t^3}
{\Gamma_t^{(0)}\over \Gamma_f^{(0)}},
\label{LW1}
\end{equation}
\begin{equation}
p_f^{(0)}+p_t^{(0)}=1.
\label{LW2}
\end{equation}
We see that, regardless of the initial conditions, the probability
distribution rapidly approaches the stationary distribution
(\ref{LW1}).  

Let us compare the distribution (\ref{LW1}) with that
discussed by Lee and Weinberg \cite{LW}. The distributions agree provided 
that the nucleation rates which appear in their expressions are taken as
$\Gamma_{t/f}^{(0)}$.
Lee and Weiberg also argue that 
the ratio (\ref{LW1}) can be given to one loop order as
\begin{equation}
{\kappa_t\over \kappa_f}= 
{H_f^3\over H_t^3} \exp[-(3/8)(\rho_t^{-1}-\rho_f^{-1})],
\label{ratio}
\end{equation}
The absence of determinantal prefactors in this expression 
is justified by the fact that
the bounce solution for true and false vacuum decay are the same, hence
the primed determinants corresponding to fluctuations around the bounce
cancel out in the ratio of rates. The effect of determinants corresponding 
to fluctuations around the true and false vacuum background instantons
is just to renormalize the values of the corresponding effective cosmological
constants. Hence, in (\ref{ratio}), the vacuum densities $\rho_{t/f}$ should
be understood as their `one loop corrected' values.  

In Ref. \cite{Garriga} the nucleation rates for true and
false vacuum bubbles were studied in the case when the gravitational
backreaction of the bubble is ignored, so that the background geometry was
taken to be an exact de Sitter space with the same Hubble constant $H$
inside and outside te bubbles. It was found that the number distribution 
of true or false ($t/f$) vacuum bubbles centered around the point 
$(t_0,{\bf x})$ could be written as 
$d{\cal N}_{t/f}=|\lambda_{t/f}|\exp{(3Ht_0)}\ d^3 {\bf x}\  dt_0$. Here 
$\lambda_{t/f}=A \exp{(-B_{t/f})}$, where $A$ is a primed determinant 
which is the same for true and false vacuum bubbles,
and $B_{t/f}=S_B-S_{t/f}$ 
is the difference between the bounce action and the background Euclidean
action. The expression for $d{\cal N}$ is proportional to the 
physical volume element at time $t_0$, given by 
$\exp{(3Ht_0)}\ d^3 {\bf x}$, so
we can identify $|\lambda_{t/f}|$ with
the rates $\Gamma^{(0)}_{t/f}$ defined above. Therefore, 
we have 
$(\Gamma^{(0)}_{t}/\Gamma^{(0)}_{f}) = \exp[S_t-S_f]$,
in agreement with Eqs. (\ref{LW1},\ref{ratio}).

We introduced $p_f$ and $p_t$ as fractions of co-moving observers in
false and true vacuum, respectively.  When the stationary distribution
(\ref{LW1}) is reached, an alternative interpretation will also be
useful.  The world line of each observer will repeatedly cross between
true and false vacuum regions, and the quantity $p_f^{(0)}$
($p_t^{(0)}$) gives the fraction of proper time the observer spends in
false (true) vacuum.

Instead of the proper time $\tau$, one can use some other time
variable, $t$, along the observer's worldlines.  A possible choice is
\begin{equation}
dt=H^\alpha(\tau)d\tau,
\label{t}
\end{equation}
with $\alpha=0$ corresponding to the proper time and $\alpha=1$ to the
``scale factor time''.  For an arbitrary $\alpha$,
the evolution equations still have the form (\ref{eqpf}),(\ref{eqpt}),
with $\tau$ replaced by $t$ and
\begin{equation}
\kappa_f=\Gamma_f^{(0)}{4\pi\over{3}}H_f^{-\alpha-3},
\label{kappafa}
\end{equation}
\begin{equation}
\kappa_t=\Gamma_t^{(0)}{4\pi\over{3}}H_t^{-\alpha-3}.
\label{kappata}
\end{equation} 
The stationary solution now is
\begin{equation}
{p_f^{(\alpha)}\over{p_t^{(\alpha)}}}=\left({H_f\over{H_t}}
\right)^\alpha {p_f^{(0)}\over{p_t^{(0)}}}.
\label{LWa}
\end{equation}
The $\alpha$-dependence of (\ref{LWa}) can be easily understood: $p_f$
($p_t$) is proportional to the amount of time spent by a co-moving
observer in false (true) vacuum, and if the time variable is changed
as in (\ref{t}), the ratio $p_f/p_t$ is modified by a factor
$(H_f/H_t)^\alpha$. 

\subsection{A more realistic model}

Let us now consider a two-field model of the type (\ref{LM}), except we
shall assume that the false vacuum at $\chi=0$ corresponds to a single
point, rather than a flat direction, in the field space.  This is the
case, for example, in models where $\chi$ and $\phi$ represent the
radial and angular parts of a single complex field, $\Phi=\chi
e^{i\phi}$.  We shall assume further that the effective potential for
the field $\phi$ is of the ``new'' inflation type and has a slow roll
region $\phi_*^{(1)}<\phi<\phi_*^{(2)}$.  Finally, to simplify the
discussion, we shall disregard the evolution between the end of slow
roll and true vacuum domination.  That is, we shall assume that when
the field $\phi$ rolls down to $\phi_*^{(j)}$, it gets directly to the
true minimum of the potential with energy density $\rho_t^{(j)}$,
where $j=1,2$.  We shall refer to $\phi_*^{(j)}$ as ``thermalization
points'' and to the corresponding minima of the potential as the first
and the second true vacua, $TV(1)$ and $TV(2)$. (If $\rho_t^{(1)}\not=
\rho_t^{(2)}$, then of course only one of these vacua is a truly true
vacuum).  The false vacuum will be abbreviated as $FV$.

Once again we introduce an ensemble of co-moving observers and define
$p_{t1}(t)$, $p_{t2}(t)$ and $p_f(t)$ to be the fractions of
the observers in $TV(1)$, $TV(2)$ and $FV$, respectively.  We also define
${\cal P}(\phi,t)d\phi$ as the fraction of observers who are, at
time $t$, located in 
slow roll regions with the inflaton field between $\phi$ and
$\phi+d\phi$.  
We can now combine the analysis in the preceeding subsection with the
standard formalism of stochastic inflation to obtain the system of
equations describing the evolution of our model:
\begin{equation}
{\partial{\cal P}\over{\partial t}}=-{\partial J\over{\partial\phi}}
-{\tilde\kappa}_f(\phi){\cal P}+\kappa_f(\phi) p_f,
\label{eq1}
\end{equation}
\begin{equation}
{dp_{t1}\over{dt}}=-\kappa_{t1}p_{t1}+\kappa_{f1}p_f -J_1,
\label{eq2}
\end{equation}
\begin{equation}
{dp_{t2}\over{dt}}=-\kappa_{t2}p_{t2}+\kappa_{f2}p_f +J_2,
\label{eq3}
\end{equation}
\begin{equation}
{dp_{f}\over{dt}}=-(\kappa_{f1}+\kappa_{f2})p_f +\kappa_{t1}p_{t1}
+\kappa_{t2}p_{t2} -p_f\int \kappa_f(\phi)d\phi +\int{\tilde\kappa}_f
(\phi){\cal P}d\phi.
\label{eq4}
\end{equation}
Here, $\kappa_{fj}$ corresponds to tunneling from $FV$ to $TV(j)$,
$\kappa_{tj}$ to tunneling from $TV(j)$ to $FV$; they are
given by Eqs.(\ref{kappafa}), (\ref{kappata}) with an extra index $j$
added to the appropriate quantities.  $\kappa_f(\phi)d\phi$
corresponds to
tunneling from $FV$ to a value $\phi$ in the interval $d\phi$ in the 
slow roll region, and ${\tilde\kappa}_f(\phi)$ to tunneling from a
slow roll region with a given value of $\phi$ to $FV$.  By analogy
with (\ref{kappafa}),(\ref{kappata}) we can write
\begin{equation}
\kappa_f(\phi)=\Gamma_f^{(0)}(\phi){4\pi\over{3}}H_f^{-\alpha-3},
\label{kappafphi}
\end{equation}
\begin{equation}
{\tilde\kappa}_f(\phi)={\tilde\Gamma}_f^{(0)}(\phi)
{4\pi\over{3}}H^{-\alpha-3}(\phi),
\label{tildekappafphi}
\end{equation}
To simplify the equations, we have disregarded tunneling between
$TV(1)$ and $TV(2)$ and between $TV(j)$ and the slow roll region.
These effects can be easily included if necessary.

The current $J(\phi,t)$ in Eq.(\ref{eq1}) is given by
\begin{equation}
J(\phi,t)=-D^{1-\beta}(\phi){\partial\over{\partial\phi}}
[D^\beta(\phi){\cal P}(\phi,t)]+v(\phi){\cal P}(\phi,t),
\label{J}
\end{equation}
where the first term on the right-hand side describes the 
``diffusion'' of
the field $\phi$ due to quantum fluctuations, with a diffusion
coefficient
\begin{equation}
D(\phi)=H^{3-\alpha}(\phi)/8\pi^2,
\label{D}
\end{equation}
the second term describes the classical ``drift'' with velocity
\begin{equation}
v(\phi)=-H^{-\alpha}(\phi)H'(\phi)/4\pi,
\label{v}
\end{equation}
and
\begin{equation}
H^2(\phi)=8\pi V(\phi)/3.
\end{equation}
The integration in Eq.(\ref{eq4}) is from $\phi_*^{(1)}$ to
$\phi_*^{(2)}$, and the quantities $J_j$ in
Eqs.(\ref{eq2}),(\ref{eq3}) are defined as $J_j(t)\equiv
J(\phi_*^{(j)},t)$.  The normalization condition
\begin{equation}
\int {\cal P}d\phi +p_f+p_{t1}+p_{t2}=1
\label{normc}
\end{equation}
is preserved by the evolution equations (\ref{eq1})-(\ref{eq4}).

The parameter $\beta$ in Eq.(\ref{J}) for the current represents the
factor-ordering ambiguity in the diffusion equation with a
position-dependent diffusion coefficient.  The choices $\beta=1/2$ and
$\beta=1$ are usually referred to as Stratonovich and Ito factor ordering,
respectively.  

The boundary conditions for Eq.(\ref{eq1}) are
\begin{equation}
{\partial\over{\partial \phi}}[D^\beta(\phi){\cal P}(\phi,t)]_{\phi
=\phi_*^{(j)}}=0.
\label{bc}
\end{equation}
They ensure that, once the field $\phi$ rolls down to $\phi_*^{(j)}$,
it does not diffuse back to the slow-roll regime, but rather stays in
the true vacuum $TV(j)$ until it tunnels to $FV$.

The system of equations (\ref{eq1})-(\ref{eq4}) can be written
symbolically in the operator form,
\begin{equation}
{d{\bf P}\over{dt}}={\bf M}{\bf P},
\label{mateq}
\end{equation}
where the ``vector'' ${\bf P}(t)$ is ${\bf P}=\{ p_f,p_{t1},p_{t2},
{\cal P}(\phi)\}$.  With an appropriate discretization of $\phi$, this
can be rewritten in the form of a ``master equation'',
\begin{equation}
{dP_i\over{dt}}=\sum_j (w_{ij}P_j -w_{ji}P_i)\equiv \sum_jM_{ij}P_j.
\label{mastereq}
\end{equation}
Each quantity $w_{ij}$ is positive and has the meaning of the
transition rate from state $j$ to state $i$.  
The matrix $M_{ij}$ can be represented as
\begin{equation}
M_{ij}=w_{ij}-\delta_{ij}\sum_k w_{ki}
\end{equation}
and has the properties
\begin{equation}
M_{ij}\geq 0 ~~~ (i\not= j),
\label{prop1}
\end{equation}
\begin{equation}
\sum_i M_{ij}=0.
\label{prop2}
\end{equation}
The latter property ensures the conservation of probability, $\sum_i
{\dot P_i}=0$.  It also indicates that the matrix ${\bf M}$ has a left
eigenvector ${\bf Q}=\{ 1,1,1, ...\}$ with zero eigenvalue, ${\bf
QM}=0$.  Since ${\bf M}$ and its transpose have the same eigenvalues,
it follows that ${\bf M}$ should also have a right zero eigenvector,
\begin{equation}
{\bf MP_0}=0,
\end{equation}
indicating that our system of equations has a stationary solution.

The familiar method of solving Eq.(\ref{mastereq}) using a
decomposition in eigenvectors cannot, in general, be applied because
the matrix ${\bf M}$ is not generally symmetric.  However, some
properties of its solutions can be derived using only
Eqs.(\ref{prop1}),(\ref{prop2}).  We shall assume that the
matrix ${\bf M}$ is irreducible (otherwise, the master equation
(\ref{mateq}) describes several independent processes which can each
be described by a master equation with an irreducible matrix ${\bf
M}$).  Then it can be shown \cite{vanKampen}
that (i) the zero eigenvalue, $\gamma_0=0$, is
non-degenerate, (ii) all components of the corresponding eigenvector
${\bf P_0}$ are non-negative, (iii) all other eigenvalues of ${\bf M}$
satisfy ${\rm Re}\gamma_n <0$, and (iv) the asymptotic behavior of
${\bf P}(t)$ at late times is
\begin{equation}
{\bf P}(t\to\infty)={\bf P}_0.
\end{equation}    
In other words, the stationary solution is unique, and all solutions
approach this stationary solution at $t\to\infty$.
Although these results have been rigorously derived only for a finite set of
$P_n$, we shall assume that they are still valid in the continuum
limit. 

If all eigenvalues of ${\bf M}$ are non-degenerate, then ${\bf M}$ can
be diagonalized, and the general solution of (\ref{mateq}) can be
written as
\begin{equation}
{\bf P}(t)= \sum_{n=0}^\infty {\bf P}_n e^{\gamma_n t},
\label{matsol}
\end{equation}
where ${\bf P}_n$ are eigenvectors of the operator ${\bf M}$,
and $\gamma_n$ are the corresponding eigenvalues,
\begin{equation}
{\bf M}{\bf P}_n =\gamma_n {\bf P}_n.
\label{eigen}
\end{equation}
Since ${\bf M}$ is real, its eigenvalues and eigenvectors come in
complex conjugate pairs.  In the case of degenerate eigenvalues, the
solution is more complicated \cite{Petrovsky}.

In the absence of recycling, $\kappa_{tj}={\tilde \kappa}_f(\phi)=0$,
and the stationary solution of the system (\ref{eq1})-(\ref{eq4}) is
trivial: $p_{tj}=const, p_f={\cal P}(\phi)=0$.  The standard analysis
of stochastic inflation \cite{AV83,Starob86,LLM94} 
has been done for a slow-roll
inflation without a metastable false vacuum.  Then Eq.(\ref{eq1})  
reduces to a Fokker-Planck equation for  ${\cal P}(\phi,t)$,
\begin{equation}
{\partial{\cal P}\over{\partial t}}=-{\partial J\over{\partial\phi}}
\equiv {\cal MP}
\label{FPeq}
\end{equation}
It can be shown \cite{Risken} that, with an appropriate choice of a
scalar product, the differential operator ${\cal M}$
is Hermitian.  Hence, all its eigenvalues are real and
the eigenvectors form a complete orthonormal set.  An eigenvector
expansion of the form (\ref{matsol}) is then always possible, and the
asymptotic behavior of ${\cal P}(t)$ is 
\begin{equation}
{\cal P}(\phi,t\to\infty)= f(\phi) \exp (\gamma_m t).
\label{calp}
\end{equation}    
Here, $\gamma_m<0$ is the largest eigenvalue of ${\cal M}$.
In this standard approach the distribution 
${\cal P}(\phi,t)$ is not normalized: the observers who left the slow
roll range through the boundaries at $\phi_*^{(j)}$, never return, and
$\int {\cal P}d\phi$ decreases exponentially with time.

\subsection{The fractal dimension}

In the case of ``new'' inflation without recycling, the inflating part
of the universe represents a self-similar fractal of dimension
\cite{Aryal,AV95} 
\begin{equation}
d = 3-\gamma_m H_m^{\alpha -1}.
\label{fracdim}
\end{equation}
Here, as in Eq.(\ref{calp}), $\gamma_m$ is the largest eigenvalue of
the Fokker-Planck operator ${\cal M}$, and $H_m$ is the expansion rate
at the maximum of $V(\phi)$ \cite{Foot3}.  For a co-moving sphere of
radius $R$ centered on a point in the inflating region, the inflating
volume within the sphere is ${\cal V}(R)\propto R^d$, which is a
fraction
\begin{equation}
f(R)\propto R^{d-3}
\end{equation}
of the total volume of the sphere.  As the sphere expands, $R\propto
\exp (H_m^{1-\alpha}t)$, this fraction decreases as $f\propto\exp
(\gamma_m t)$, and vanishes at $t\to\infty$.  Hence, the inflating
region represents a set of measure zero in the limit $t\to\infty$.

On the other hand, the inflating part of the volume in a recycling
universe is constantly replenished by tunneling from the true vacuum.
As a result, the inflating region occupies a non-vanishing fraction of
the total volume, so that ${\cal V}(R)\propto R^3$ and $d=3$.

We note, however, that a recycling universe does contain fractal
regions of dimension $d<3$.  Take for example the Lee-Weinberg model
of Section A and consider a co-moving volume which is initially filled
with $FV$.  What remains of this $FV$ in the limit $t\to\infty$ is a
fractal of dimension \cite{AV92}
\begin{equation}
d_f=3-\Gamma_f^{(0)}{4\pi\over{3}}H_f^{-4}.
\end{equation}
All the remaining part of the volume is occupied by true vacuum
bubbles, but what remains of the $TV$ inside the bubbles at
$t\to\infty$ is also a fractal of dimension
\begin{equation}
d_t=3-\Gamma_t^{(0)}{4\pi\over{3}}H_t^{-4}.
\end{equation}
The $FV$ bubbles inside each $TV$ bubble have dimension $d_f$, and
they are in turn filled by $TV$ bubbles of dimension $d_t$.  
The fractal structure of realistic models is of course more
complicated. 

\subsection{Choosing the factor ordering}

One of the problems with interpreting the results of the stochasic
inflation formalism is the dependence of these results on the choice
of the time variable $t$ and on the factor ordering in the
Fokker-Planck equation (\ref{eq1}), (\ref{J}) \cite{LLM94,WV96}.  
We have parametrized
these choices by the parameters $\alpha$ and $\beta$.  Now we are
going to argue that there is a preferred choice of $\beta$ which
allows at least a partial resolution of these problems.

As we discussed in Section IV.A, the stationary distribution ${\bf
P}_0$ gives the fraction of time spent by a co-moving observer in
false and true vacua and in different parts of the slow-roll range.
This distribution should of course depend on how we define the time
variable, but the dependence should be trivial,
\begin{equation}
{\cal P}(\phi)\propto H^\alpha(\phi), ~~~~~~ p_f\propto H_f^\alpha,
~~~~~~ p_{tj}\propto H_{tj}^\alpha.
\label{alphadep}
\end{equation}
To ensure that this is indeed the case, we can require that in the
stationary version of Eq.(\ref{mateq}),
\begin{equation}
{\bf M}{\bf P}=0,
\end{equation}
all components of ${\bf P}$ appear only in combinations
$H^{-\alpha}(\phi) {\cal P}(\phi)$, $H_f^{-\alpha}p_f$,
$H_{tj}^{-\alpha} p_{tj}$, and that there is no other dependence on
$\alpha$.  This fixes
\begin{equation}
\beta=1,
\end{equation}
which corresponds to Ito factor ordering.

It should be noted that the family of factor ordering choices
parametrized by $\beta$ in Eq.(\ref{J}) does not include all
possibilities.  Although our requirement (\ref{alphadep}) is
sufficient to fix $\beta$ uniquely, it does not determine a unique
factor ordering in the general case.  For example, we could replace
the diffusion term in (\ref{J}) by
\begin{equation}
D^{1-\beta}(\phi){\partial\over{\partial\phi}}[D^\beta(\phi) {\cal
P}(\phi,t)] \to h^{-1}(\phi){\partial\over{\partial\phi}}[h(\phi)
D(\phi){\cal P}(\phi,t)].
\end{equation}
The condition (\ref{alphadep}) is satisfied for an arbitrary,
$\alpha$-independent function $h(\phi)$.

\subsection{Some solutions}

The system of equations (\ref{eq1})-(\ref{eq4}) looks rather
intimidating, but stationary solutions of this system can actually be
found in some special cases.  

Let us first assume that tunneling from
slow roll to false vacuum can be neglected, that is, ${\tilde\kappa}_f
(\phi)=0$.  This means that the evolution proceeds along the path
$FV\to$ slow roll $\to TV\to FV \to ...$, possibly with occasional
tunneling directly from $FV$ to $TV$.  Then, with $\beta=1$, the 
stationary version of (\ref{eq1}) can be written as
\begin{equation}
\partial_\phi J(\phi)=\kappa(\phi)p_f,
\end{equation}
where
\begin{equation}
J(\phi)=-\partial_\phi [D(\phi){\cal P}(\phi)]+v(\phi){\cal P}(\phi).
\end{equation}
This is easily integrated to give
\begin{equation}
{\cal P}(\phi)=-{8\pi^2 p_f\over{H^{3-\alpha}(\phi)}}e^{\pi/H^2(\phi)}
\left[ \int_{\phi_*^{(1)}}^\phi d\phi' e^{-\pi/H^2(\phi')}
\int_{\phi_*^{(1)}}^{\phi'} d\phi'' \kappa_f(\phi'')
+C_1\int_{\phi_*^{(1)}}^\phi d\phi' e^{-\pi/H^2(\phi')} +C_2\right],  
\label{sol1}
\end{equation}
where we have used the expressions (\ref{D}),(\ref{v}) for $D(\phi)$
and $v(\phi)$.  

The integration constants $C_1$, $C_2$ can now be found from the
boundary conditions (\ref{bc}) (with $\beta=1$).  Thus we obtain the
distribution ${\cal P}(\phi)$ in terms of $p_f$.  
Evaluating $J_j=J(\phi_*^{(j)})$ and substituting in
Eqs.(\ref{eq2}),(\ref{eq3}), we find $p_{tj}$ in terms of $p_f$.
Finally, $p_f$ is found from the normalization condition (\ref{normc}).
[Note that $p_f$ cannot be found from Eq.(\ref{eq4}) which is a linear
combination of the preceeding three equations (\ref{eq1})-(\ref{eq3})].
The resulting expressions are rather cumbersome and we shall not
reproduce them here.  

As another example, we take a potential $V(\phi)$ of the form
considered in Ref. \cite{AV95}, which consists of a flat portion where 
$H(\phi)=const,~ \kappa_f(\phi)=const,~ {\tilde\kappa}_f(\phi)=const$,
surrounded by two regions with a relatively large slope where the
diffusion term is negligible.  In the flat range of $\phi$, the
Fokker-Planck equation is trivially solved.  In the diffusionless
regions, 
\begin{equation}
\partial_\phi [v(\phi){\cal P}(\phi)]+{\tilde\kappa}_f(\phi){\cal
P}(\phi)=\kappa_f(\phi)p_f,
\end{equation}
and a straightforward integration gives
\begin{equation}
{\cal P}(\phi)=-4\pi p_f{H^\alpha(\phi)\over{H'(\phi)}}
e^{-g(\phi)}\left[ \int_{\phi_*^{(1)}}^\phi d\phi'
\kappa_f(\phi')e^{g(\phi')}+C_1 \right],
\label{nodiff}
\end{equation}
where
\begin{equation}
g(\phi)=\int_{\phi_*^{(1)}}^\phi d\phi'{{\tilde\kappa}_f(\phi')
\over{v(\phi')}},
\end{equation}
for the region bounded by $\phi_*^{(1)}$, and similarly for the second
region bounded by $\phi_*^{(2)}$.
The integration constants and the values of $p_f$ and $p_{tj}$ can be
determined by matching the solutions at the boundaries between the
flat and diffusionless regions and by using
Eqs.(\ref{eq2}),(\ref{eq3}) and the normalization condition
(\ref{normc}).

In the general case, the solution (\ref{nodiff}) should still apply
in the range of $\phi$ 
sufficiently close to the thermalization points, where diffusion is
negligible.  If tunneling between this range and the false vacuum is
unimportant, then the solution takes a particularly simple form,
\begin{equation}
{\cal P}(\phi)=-4\pi C_1 p_f{H^\alpha(\phi)\over{H'(\phi)}},
\label{nodiff'}
\end{equation}
and similarly for the range of $\phi$ near $\phi_*^{(2)}$.  The
constants $C_1$ and $C_2$ can be determined only after solving the
equation in the entire range of $\phi$.

The distribution (\ref{nodiff'}) in the diffusionless
range of $\phi$ should be compared with the corresponding distribution
in the absence of recyclings \cite{AV95},
\begin{equation}
{\cal P}(\phi)=C{H^\alpha(\phi)\over{H'(\phi)}}\exp\left[ -4\pi\gamma_m
\int_{\phi_*^{(1)}}^\phi d\phi'{H^\alpha(\phi')\over{H'(\phi')}}\right].
\label{nodiff''}
\end{equation}
This can be
drastically different from (\ref{nodiff'}) even if the tunneling
probabilities are very small.  Hence, there is no continuous transition
between recycling and no-recycling regimes in the limit of vanishing
tunneling probabilities.

\subsection{Physical volume distribution}

The function ${\bf P}(t)$ characterizes the distribution of co-moving
volume between false and true vacua and different values of $\phi$ in
the slow roll regime.  One can introduce a similar function for the
physical volume distribution \cite{GLM87,NS89,Mijic},
\begin{equation}
{\bf {\tilde P}}(t)=\{{\tilde{\cal P}}(\phi,t),{\tilde p}_f(t), {\tilde
p}_{t1}(t),{\tilde p}_{t2}(t)\}.
\end{equation}
It is defined so that ${\tilde {\cal P}}(\phi,t)d\phi$ is the physical
volume occupied by slow roll regions with $\phi$ in the range $d\phi$
at time $t$, {\it etc}.  The distribution ${\bf {\tilde P}}(t)$
satisfies a modified master equation,
\begin{equation}
{d{\bf {\tilde P}}\over{dt}}={\bf M{\tilde P}}+3{\bf
H}^{1-\alpha}{\bf {\tilde P}}\equiv {\bf{\tilde M}{\tilde P}},
\label{eqptilde}
\end{equation}
where the operator ${\bf H}^{1-\alpha}$ is given by
\begin{equation}
{\bf H}^{1-\alpha}=diag\{H^{1-\alpha}(\phi),H_f^{1-\alpha}, 
H_{t1}^{1-\alpha},H_{t2}^{1-\alpha}\}.
\end{equation}
In an infinite universe, the volume is of course infinite, but the
distribution ${\bf {\tilde P}}(t)$ can be defined on a fixed co-moving
part of the universe.  The form of the distribution at large $t$ is
independent of the choice of the co-moving region.

In the discretized version of Eq.(\ref{eqptilde}), the matrix
${\tilde{\bf M}}$ does not have the property (\ref{prop2}), and the
standard theorems for the master equation do not apply.  However, the
following statements can still be proved \cite{W} using the
Perron-Frobenius theorem about non-negative matrices:  (i)
${\tilde{\bf M}}$ has a real eigenvalue ${\tilde \gamma}_0$ which is
greater than the real parts of all other eigenvalues and which is
bounded by
\begin{equation}
3H_{min}^{1-\alpha}\leq {\tilde \gamma}_0\leq 3H_{max}^{1-\alpha},
\end{equation}
where $H_{max}$ and $H_{min}$ are respectively the largest and the
smallest values of $H$;
(ii) the corresponding eigenvector ${\tilde{\bf P}}_0$ has
non-negative components; (iii) ${\tilde \gamma}_0$ is non-degenerate if
${\tilde{\bf M}}$ is irreducible.  
The late-time asymptotic behavior of ${\tilde{\bf P}}(t)$ is
\begin{equation}
{\bf {\tilde P}}(t\to\infty)={\bf {\tilde P}}_0 e^{{\tilde\gamma}_0
t},
\end{equation}

In contrast to the co-moving distribution ${\bf P}(t)$, the physical
volume distribution ${\bf {\tilde P}}(t)$ has a sensitive dependence
on the choice of the time paramater $\alpha$, which does not reduce to
the trivial form (\ref{alphadep}) \cite{LLM94,AV95,WV96}.  
The equation (\ref{eqptilde}) for
${\bf {\tilde P}}(t)$ is simplified if we choose the scale factor
time, $\alpha=1$.  In this case, ${\bf H}^{1-\alpha}=1$, and the
solutions of (\ref{eqptilde}) and (\ref{mateq}) are related by
\cite{LLM94}. 
\begin{equation}
{\bf {\tilde P}}_{\alpha=1}(t)=e^{3t}{\bf P}_{\alpha=1}(t).
\end{equation}

\subsection{Discussion}

The main conclusion of our analysis in this Section is that the
distribution of co-moving observers in a recycling universe rapidly
approaches a stationary form.  This asymptotic distribution can be
obtained as the eigenvector of the ``master'' operator ${\bf M}$
with a zero eigenvalue, $\gamma_0=0$:
\begin{equation}
{\bf M}{\bf P}_0=0.
\end{equation}

The formalism we developed here can be straightforwadly extended to
include radiation and matter dominated periods between thermalization
and true vacuum domination.  One expects to find that the asymptotic
distribution will still be stationary, with fixed fractions of
co-moving volume occupied by radiation and matter-dominated regions.  

The picture in which co-moving ``observers'' move in an endless cycle
between $FV$, slow roll, matter domination, and $TV$, may be
oversimplified.  It should be understood, of course, that no material
observer is likely to survive the transition between $TV$ and $FV$.
Even if we think of an ``observer'' as an indestructible test
particle, there seems to be no unique way to continue his world line
into a nucleating bubble, since the surface at which we glue the bottom of
the false vacuum bubble onto the true vacuum can be chosen in different ways.
So we should probably think of our
``observers'' as being smeared over a horizon-size volume.  

Next, we note that density fluctuations produced during inflation (or
generated by topological defects) result in the formation of bound
objects during the matter-dominated era.  Some of these objects
collapse to black holes, and observers in matter-dominated regions
have a finite probability (per unit time) to end their world lines at
black hole singularities. However, black holes eventually evaporate,
giving back their volume to the true vacuum. Hence this effect would
not alter our conclusions. The same happens with black holes that
may spontaneously nucleate in false or true vacuum \cite{bhn}. The
rate of black hole pair production in true vacuum is proportional
to $\exp[-1/8\rho_t]$. This rate is considerably larger than that
for nucleation of a false vacuum bubble, which in the case 
$\rho_t<<\rho_f$ is 
proportional  $\exp[-3/8\rho_t]$. It is also possible that nucleated black
holes may act as seeds for false vaccum bubble nucleation, as they do
for true vacuum bubbles \cite{SH}.

Finally, the inflaton potential $V(\phi)$
can be of the ``chaotic'' inflation type, with the slow roll range of
$\phi$ extending to Planckian energy densities.  Then there is a
finite probability for an observer to get into this Planckian domain,
where the classical description of spacetime breaks down.  In the
stochastic inflation formalism, this is accounted for by introducing a
``Planck boundary'' at some $\phi=\phi_p$, such that $V(\phi_p)\sim1$.
The loss of observers through the Planck boundary
will generally result in $\gamma_0>0$ and $d<3$. 

The same phenomenon of 
loss of observers will also occur if some of the minima of the potential
have vanishing or negative cosmological constant. 
Once some co-moving 
volume falls into one of these vacua it has no chance of being recycled.

\section{Problems with predictions}

Different thermalized regions of the universe are generally
characterized by different values of the constants of Nature and of
the cosmological parameters (such as the density parameter $\Omega$).
In the model that we used as an example in Section V, the universe can
thermalize into two types of vacua, $TV1$ and $TV2$, and thus we have
two possible sets of constants of Nature.  The number of possibilities
can, in principle, be much larger, and in some models the
``constants'' can even take values in a continuous range (examples are
the effective gravitational constant in a Brans-Dicke-type theory
\cite{GBL95} and
the density parameter in some models of ``open'' inflation \cite{LM95})
An intriguing question is whether or not we can ``predict'' which set
of the constants we are most likely to observe.

One can try to determine the probability distribution for the
constants with the aid of the ``principle of mediocrity'' which
asserts that we are ``typical'' among the civilizations inhabiting the
universe \cite{AV94,Foot5,Albrecht,GBL95}.  Here, the universe is
understood as the entire spacetime; our civilization is assumed to be
typical among all civilizations, including those that no longer exist
and those that will appear in the future.  The probability for us to
observe a given set of constants is then proportional to the total
number of civilizations in the corresponding type of thermalized
regions.  This number can be represented as the number of galaxies
(which one can hope to estimate) times the number of civilizations per
galaxy (which is left undetermined until the evolution of life and 
consciousness are better understood).  Some of the constants, such as
the cosmological constant or the density parameter, are not expected
to affect the chances for a civilization to develop in a given galaxy, 
so one can hope to determine the probability distribution for such 
constants without any biological input.

In the case of a closed universe and 
finite (non-eternal) inflation, this prescription for
calculating probabilities is unambiguous.  If the universe is
spatially infinite, one can simply use the prescription for a fixed
(sufficiently large) co-moving volume.  However, in an eternally
inflating universe the spacetime volume and the number of
civilizations are infinite, even for a region of a finite co-moving
size.  One can deal with this problem by simply introducing a time
cutoff and counting only the number of civilizations $N_j(t_c)$ 
that appeared prior to some
moment of time $t_c$.  Here, the index $j$ refers to the type of
thermalized region.  The ratio of probabilities 
can then be defined as the limit
\begin{equation}
{p_1\over{p_2}}= {\rm lim}_{t_c\to\infty}{N_1(t_c)\over{N_2(t_c)}}.
\end{equation}
One finds, however, that the resulting probability distribution is
extremely sensitive to the choice of the time variable $t$
\cite{Foot6,LLM94}.  This gauge-dependence casts doubt on any
conclusions reached using this approach.

An alternative procedure, suggested in \cite{AV95}, is to introduce a
cutoff at the time $t_c^{(j)}$, when all but a small fraction
$\epsilon$ of the co-moving volume destined to thermalize into regions
of type $j$ has thermalized.  The value of $\epsilon$ is taken to be the same
for all types of thermalized regions, but the corresponding cutoff
times $t_c^{(j)}$ are generally different.  The limit $\epsilon\to 0$
is taken after calculating the probability distribution for the
constants.  It was shown in \cite{AV95,WV96} that the resulting
probabilities are essentially insensitive to the choice of time
parametrization.  However, the same problem appears in a different
guise.  Linde and Mezhlumian \cite{LM96} have found a family of
gauge-invariant cutoff 
procedures parametrized by a dimensionless parameter
$q$, with $q=0$ corresponding to the $\epsilon$-procedure described
above.  This indicates that the invariance requirement alone is not
sufficient to define the probabilities uniquely.  Some additional
requirements that can fix the parameter $q$ have been discussed in
\cite{WV96}.

Now, recycling introduces one more difficulty.  In the absence of
recycling, co-moving regions could be uniquely characterized by the
type of thermalized region they will evolve into.  But in a recycling
universe each co-moving region goes through an endless succession of
different types of thermalization.  Hence, the $\epsilon$-procedure
cannot be implemented in its present form \cite{Foot7}.

In the face of these difficulties, one could look for entirely
different approaches to defining the probabilities.  One possibility
is to abandon the requirement of gauge-invariance and assert that
there is a preferred choice of the time variable $t$.  If this
approach is taken, then there is, arguably, a good reason to take the
scale-factor time, $t=\ln a$, as the preferred choice \cite{Starob86,SB91}.
The only variables that can be used as clocks in an inflating universe
are the inflaton field $\phi$ and the scale factor $a$.  The main
requirement for a clock is a predictable classical behavior.  In the range
of $\phi$ where quantum fluctuations are important, $\phi$ is not
suitable for this role, 
and the only remaining variable to be used as a clock is $a$.

Another possibility is to abandon the principle of mediocrity and invoke
the ideal observers that we used to define the distribution ${\bf
P}(t)$, rather than physical observers, to calculate probabilities.
In a recycling universe, the worldline of an ideal observer crosses an
infinite number of inflating and thermalized regions.  The
probabilities for different types of thermalized regions can then be
defined as relative frequencies at which these regions are encountered
along the worldline.  This definition is obviously gauge-invariant.
In the model of Section V.B, it gives
\begin{equation}
p_1/p_2=|J_1/J_2|.
\label{eternal}
\end{equation}
The gauge-invariance of (\ref{eternal}) is easily verified from
Eqs.(\ref{J}),(\ref{v}),(\ref{bc}),(\ref{alphadep}).
In this approach, the most probable thermalized regions may turn out
to be unsuitable for life, but this can be easily fixed by defining
appropriate conditional probabilities.

As mentioned at the end of Section III, recycling may not be complete
in models where there is a ``Planck boundary'' in the diffusion regime
or where some of the minima of the effective potential have vanishing or
negative effective cosmological constant.
In this case, the worldlines of all ideal
observers (except a set of measure zero) have a finite length, and a
natural extension of Eq.(\ref{eternal}) is
\begin{equation}
{p_1\over{p_2}}={\int_0^\infty dt |J_1(t)| \over{\int_0^\infty dt
|J_2(t)|}}. 
\label{increc}
\end{equation}
This defines the probabilities as being proportional to the total
number of encounters for a given type of region, averaged over all
observers.  The result depends on the initial distribution ${\bf
P}(0)$ at $t=0$.  Assuming that in this type of models the universe
must have a beginning, and that it can be described by quantum
cosmology, this initial distribution can be determined from the wave
function of the universe.  

Although the definition of probabilities in this approach is
gauge-invariant, it is not quite satisfactory.  The ideal
observers have very little to do with real physical observers, and it
is hard to justify why the likelihood of various observations made by
our civilization should be related to an ensemble of such ideal
observers.  

We have to conclude that, despite some effort, 
none of the approaches suggested so far appears to be particularly
compelling. 
It may turn out that, after all, an eternally inflating 
universe does not admit a uniquely defined probability distribution
for the constants of nature. If so, this does not necessarily mean that
all possible sets of constants consistent with our existence 
are equally likely. Although the ratio
$p_1/p_2$ may depend on the choice of cutoff procedure, it is
conceivable that in some cases $p_1/p_2\gg 1$ for all reasonable
choices.  We would then ``predict'' that $1$ is much more likely than
$2$.  It is possible that we will have to restrict ourselves to such
``stable'' predictions, which are insensitive to the choice of cutoff
\cite{Tegmark2}.      

\section{Conclusions}

We have shown that the picture of the superlarge-scale structure of 
the inflationary universe is significantly modified by quantum fluctuations
which bring parts of already thermalized regions back to the false vacuum,
a process which we call recycling.

In particular, the question of whether or not the Universe had a beginning
is reopened. Ignoring recycling, and under certain
rather general conditions, it has been shown in Refs. \cite{AV92,BV94} that 
inflationary models require a beginning in time. In the recycling
picture, this conclusion does not apply because it is only necessary that
any false vacuum region has a beginning in time. One can therefore
imagine a nested structure where all false vacuum regions
are just bubbles which nucleated inside  
preexisting true vacuum bubbles, which in turn nucleated inside false
vacuum bubbles and so on.  Whether or not this pattern can be
continued to the infinite past is an interesting open question.

We have extended the standard formalism of stochastic inflation
\cite{AV83,Starob86,LLM94} to study 
the probability distribution of phases in which co-moving 
observers (or co-moving volume) find themselves in the recycling universe.
Instead of the Fokker-Planck equation, one now has a more general
master equation.  We found that, in the case of complete recycling,
all solutions of this equation rapidly approach a stationary form.
This is in contrast to the standard case, where the probability
disribution decreases exponentially with time, due to the loss of
co-moving volume at thermalization.  
 
In the absence of recycling, the fractal dimension of the false vacuum 
co-moving volume at large times is lower than three. Including recycling, 
this dimension is $d=3$ because co-moving volume in false vacuum is
continually replenished by nucleating false vacuum bubbles. 
The universe ends up in a highly convoluted state, 
where the fractal dimension of any connected false or true vacuum 
region is lower than three. 

Finally, we have considered the question of ``making predictions'' for 
the constants of nature in the context of a recycling universe. 
The Principle of Mediocrity has been invoked in the past in order 
to obtain probability distributions for the constants.
For the case of finite inflation the procedure is unambiguous: 
the probability is proportional to the number of civilizations that 
observed a given set of constants 
in the entire history of the universe.  
In the case of eternal inflation, the Principle is not so easy to 
implement, because the number of thermalized regions
with given values of the constants is infinite (even in a 
finite co-moving region), and one has to introduce a
regulator. If one simply counts all civilizations below some
cut-off time, then the result depends strongly on the choice of time variable
\cite{LLM94}. A  gauge invariant cut-off prescription (i.e. one which does not 
depend on the time variable) was introduced in \cite{AV95}, but
this prescription is not unique \cite{LM96}. 
To make matters more complicated, the methods discussed so far cannot 
be directly applied to a recycling universe. We have considered
some generalizations and alternative approaches, but 
none of them is particularly compelling. 

Therefore, it seems that while the Principle of Mediocrity may offer 
some valid guidance, it may not be 
sufficient to unambiguously determine the probability distribution 
for the constants of nature. If this turns out to 
be the final answer, then ``predictions'' would only be possible in those 
cases when all reasonable cut-off methods or implementations of the Principle 
yield a similar answer.

\section*{acknowledgements}

We are grateful to Julie Traugut for suggestive conversations, to 
Serge Winitzki for very helpful discussions and to Andrei Linde for 
his comments.

This work was partially supported by NATO under grant CRG 951301.
J.G. acknowledges support from CICYT under contract AEN95-0882 and from 
European Project CI1-CT94-0004.  The work of A.V. was supported in
part by the National Science Foundation.

\section*{Figure Captions}

\begin{itemize}

\item[Fig. 1-] True vacuum bubbles (white) nucleating in false vacuum
(black). The shaded rings represent slow roll regions (external ring)
and matter or radiation dominated regions  (internal ring).

\item[Fig. 2-] Self interaction potential for the ``tunneling'' scalar
field. The energy densities in false and true vacua, denoted as
$\rho_f$ and $\rho_t$, act as an effective cosmological constant.

\item[Fig. 3-] The potential for the scalar field in the one field
model of open inflation. The sharp barrier followed by a flat plateau
gives this potential a somewhat unnatural appearance.

\item[Fig. 4-] A false vacuum bubble (black) nucleating in true vacuum
(white). Regions of slow roll and of matter and radiation domination
surrounding the bubble are indicated.

\item[Fig. 5-] The inflaton potential for the case of new inflation.

\item[Fig. 6-] De Sitter space can be viewed as a hyperboloid embedded
in a 5-dimensional Minkowski space. Here we represent a section of this
hyperboloid, along the plane $\tau, w$ in the embedding space. The
region $w+\tau>0$ of the hyperboloid can be covered with flat FRW coordinates
${\bf x}, t$. A section $w+\tau=const.$ corresponds to $t=const.$

\item[Fig. 7-] The Coleman-de Luccia instanton for bubble nucleation can 
be obtained by matching two 4-spheres of radii $H^{-1}_f$ and $H^{-1}_t$.
The two spheres join at a 3-sphere which represents the Euclideanized 
worldsheet of the domain wall.

\item[Fig. 8-] The spacetime representing a false vacuum bubble in true 
vacuum. 

\item[Fig. 9-] Conformal diagram of the false vacuum bubble spacetime,
for the Lee and Weinberg model.

\item[Fig. 10-] Same as in Fig. 9 but for the case of a one field model
of open inflation.

\end{itemize}


\begin{thebibliography}{999}

\bibitem{Review} For reviews see, for example, S.K. Blau and A.H. Guth,
in {\em 300 Years of Gravitation}, edited by S.W. Hawking and W. Israel
(Cambridge University Press, Cambridge, UK, 1978); A.D. Linde, {\em
Particle Physics and Inflationary Cosmology} (Harwood Academic, Chur, 
Switzerland, 1990); E.W. Kolb and M.S. Turner, {\em The Early Universe}
(Adison-Wesley, New York, 1990).


\bibitem{AV83} A. Vilenkin, Phys. Rev. D{\bf 27}, 2848 (1983). 

\bibitem{Linde86} A.D. Linde, Phys. Lett. B {\bf 175}, 395 (1986).

\bibitem{Aryal} M. Aryal and A. Vilenkin, 
Phys. Lett. B {\bf 199}, 351 (1987).

\bibitem{Open} J.R. Gott, Nature {\bf 295}, 304 (1982); M. Sasaki, T. Tanaka,
K. Yamamoto and J. Yokoyama, Phys. Lett B {\bf 317}, 510 (1993);
K. Yamamoto, M. Sasaki and T. Tanaka, Astrophys. J.
{\bf 455}, 412 (1995); M. Bucher, A. Goldhaber ad N. Turok, Nucl.
Phys. {\bf B}, Proc. Suppl. {\bf 43}, 173 (1995);
Phys.Rev. D {\bf 52}, 3314 (1995). 

\bibitem{LM95} A. Linde and A. Mezhlumian, Phys. Rev. D {\bf 52}, 6789 (1995).

\bibitem{HM} S.W. Hawking and I. Moss, Nucl. Phys. {\bf B224}, 180 (1983).

\bibitem{Starob86} A.A. Starobinsky, in {\em Current Topics in Field Theory,
Quantum Gravity ad Strings}, Lecture Notes in Physics, 
edited by H.J. de Vega and N. Sanchez (Springer, Heidelberg, 1986).

\bibitem{GL86} A.S. Goncharov and A.D. Linde, 
Sov. J. Part. Nucl.  {\bf 17},  369 (1986).


\bibitem{Linde91} A. Linde, Nucl. Phys. {\bf B372}, 421 (1992).


\bibitem{LW} K. Lee and E.J. Weinberg, Phys. Rev. D {\bf 36}, 1088 (1987).

\bibitem{CdL} S. Coleman and F. De Luccia, Phys. Rev. D. {\bf 21}, 3305 (1980).

\bibitem{Garriga} J. Garriga, Phys. Rev. {\bf D49}, 6327 (1994); 
{\bf D49}, 6343 (1994).

\bibitem{Linde83} A. Linde, in {\em The Very Early Universe}, edited by
G.W. Gibbons nad S.W. Hawking, Cambridge University Press (1983).

\bibitem{BV94} A. Borde and A. Vilenkin,  Phys. Rev. Lett. {\bf 72}, 
3305 (1994).


\bibitem{Borde94} A. Borde, Phys. Rev. D {\bf 50}, 3692 (1994). 

\bibitem{BVrev} A. Borde and A. Vilenkin, 
{\em Singularities in Inflationary Cosmology: A Review}, gr-qc/9612036.

\bibitem{GBL95} J. Garcia-Bellido and A. Linde, 
Phys. Rev. D {\bf 51}, 429, (1995); J. Garcia-Bellido, A. Linde and D. Linde,
Phys. Rev. D {\bf 50}, 730 (1994).

\bibitem{LLM94} A. Linde, D. Linde and A. Mezhlumian, 
Phys. Rev. D {\bf 49}, 1783 (1994).

\bibitem{AV94} A. Vilenkin, Phys. Rev. Lett. {\bf 74}, 846 (1995).




\bibitem{AV95} A. Vilenkin, Phys. Rev. D {\bf 52}, 3365 (1995). 

\bibitem{VW97} A. Vilenkin and S. Winitzki, Phys. Rev. D {\bf 55}, 548 (1997).

\bibitem{MSW97} H. Martel, P.R. Shapiro and S. Weinberg, astro-ph/9701099.

\bibitem{Foot5}  The principle of mediocrity is a version of the
anthropic principle. For a review of the latter, see e.g. B. Carter, Philos.
Trans. R. Soc. London, {\bf A310}, 347 (1983); J. D. Barrow and F. J.
Tipler, The Antropic Cosmological Principle (Clarendon Press, Oxford, 1986).
Ideas related to the principle of mediocrity have also been discussed by
Albrecht \cite{Albrecht}, Garcia-Bellido and Linde \cite{GBL95}, 
and Tegmark \cite{tegmark}.

\bibitem{Albrecht}  A. Albrecht, in {\it The Birth of the Universe and
Fundamental Forces}, ed. by F. Occhionero (Springer-Verlag, 1995).

\bibitem{tegmark} M. Tegmark, gr-qc/9704009 (unpublished).
 
\bibitem{WV96} S. Winitzki and A. Vilenkin, Phys. Rev. D {\bf 53},
4298 (1996).

\bibitem{LM96} A. Linde and A. Mezhlumian, Phys. Rev. D {\bf 53}, 4267 (1996).

\bibitem{Rama} R. Basu and A. Vilenkin, Phys. Rev D {\bf 46}, 2345 (1992);
              {\bf 50}, 7150 (1994).

\bibitem{BD} T.S. Bunch and  P.C.W. Davies, Proc. R. Soc. London, {\bf A360},
117 (1978); A. Vilenkin and L. Ford, Phys. Rev. {\bf D26}, 1231 (1982).

\bibitem{BKT} V. A. Berezin, V.A. Kuzmin and I. Tkachev, Phys. Let. B 
{\bf 120}, 91 (1983); S. Parke, Phys. Lett. B {\bf 121}, 313 (1983). 


\bibitem{FMP} W. Fishler, D. Morgan and J. Polchinski,
Phys. Rev. D {\bf 41}, 2638 (1990).

\bibitem{FGG} E. Farhi and A.H. Guth, Phys. Lett. B {\bf 183}, 149 (1987);
S.K. Blau, E.I. Guendelman and A.H. Guth, Phys Rev. D {\bf 35}, 1747 (1987). 

\bibitem{BV97} A. Borde and A. Vilenkin, {\em Violations of the Weak Energy 
Condition in Inflating Spacetimes}, gr-qc/9702019.

\bibitem{AV92} A. Vilenkin, Phys. Rev. D {\bf 46}, 2355 (1992).

\bibitem{Foot1} This is the requirement that the spacetime have a simple
causal structure. In particular, it excludes complicated topological 
interconnections between different regions of spacetime. See 
\cite{BV94,BVjap} for a precise discussion and a diagram.

\bibitem{BVjap} A. Borde and A. Vilenkin, in {\em Relativistic Astrophysics:
The Proceedings of the Eighth Yukawa Symposium}, edited by M. Sasaki,
Universal Academy Press, Japan (1995).

\bibitem{Foot2} This requires that there exist certain pairs of points
such that the spacetime volume of the difference of their pasts is finite.
See \cite{BV94,BVjap}.

\bibitem{vanKampen} N.G. van Kampen, {\it Stochastic Processes in
Physics and Chemistry} (North-Holland, Amsterdam, 1981).

\bibitem{Petrovsky} F.R. Moulton, {\it Differential Equations} (Dover,
New York, 1958).

\bibitem{Risken} H. Risken, {\it The Fokker-Planck Equation}
(Springer-Verlag, Berlin, 1989).

\bibitem{Foot3} The fractal dimension in (\ref{fracdim}) appears to
depend on the choice of the parameter $\alpha$ (that is, on the time
parametrization).  However, it has been shown in \cite{WV96} that the
product $\gamma_m H_m^{\alpha-1}$ is essentially independent of
$\alpha$.  The relative change in this product with variation of
$\alpha$ is $O(H_m^2)\ll 1$ and is comparable to its change under
variation of factor ordering.  The latter represents a genuine
uncertainty of the stochastic approach.

\bibitem{GLM87} Goncharov, Linde and Mukhanov, Int. J. Mod. Phys. A
{\bf 2}, 561 (1987).

\bibitem{NS89} Y. Nambu and M. Sasaki, Phys.Lett. {\bf B219},240
(1989).

\bibitem{Mijic} M. Mijic, Phys. Rev. {\bf D42}, 2469 (1990).

\bibitem{W} S.Winitzki, unpublished.

\bibitem{bhn} P. Ginsparg and M. Perry, Nucl. Phys. {\bf B222}, 245 (1983);
R. Bousso and S.W. Hawking,  Phys.Rev. D {\bf 54}, 6312 (1996).

\bibitem{SH} D.A. Samuel, W.A. Hiscock, Phys.Rev. {\bf D44},3052 (1991).

\bibitem{Foot6} Linde {\it et. al.} \cite{LLM94,GBL95} compared the
number of civilizations in different types of thermalized regions at a
given moment of time $t$, rather than at all times prior to $t$.  The
results of this approach are also strongly gauge-dependent.

\bibitem{Foot7} One cood choose a co-moving volume which is initially
is in the false vacuum and use the $\epsilon$-prescription to
calculate the probabilities, disregarding the recycling.  However,
such a procedure appears somewhat artificial in the case of 
a recycling universe.

\bibitem{SB91} D.S. Salopek and J.R. Bond, Phys. Rev. {\bf D43}, 1005
(1991).  

\bibitem{Tegmark2} A similar suggestion has been made independently by
Max Tegmark (private communication).

\end{thebibliography}
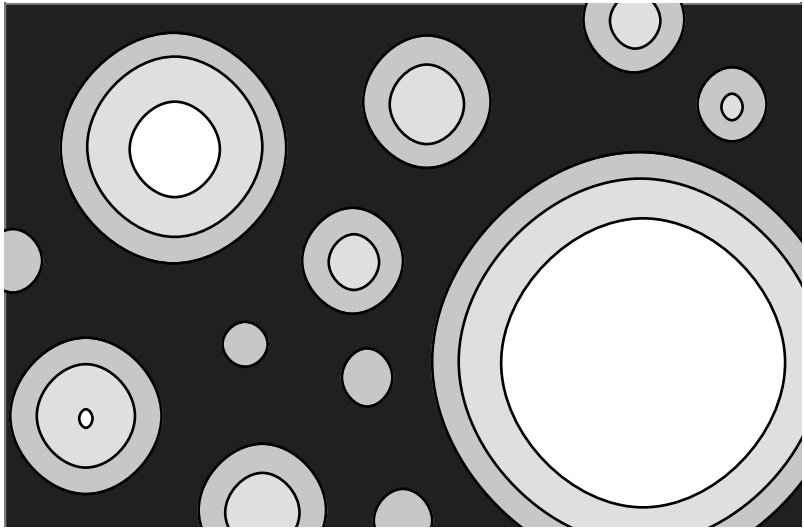
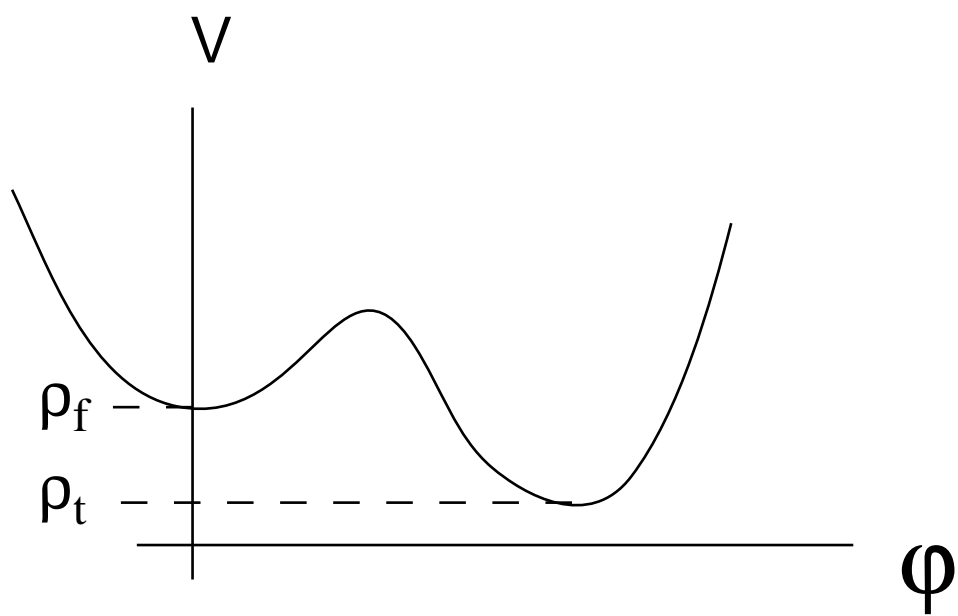
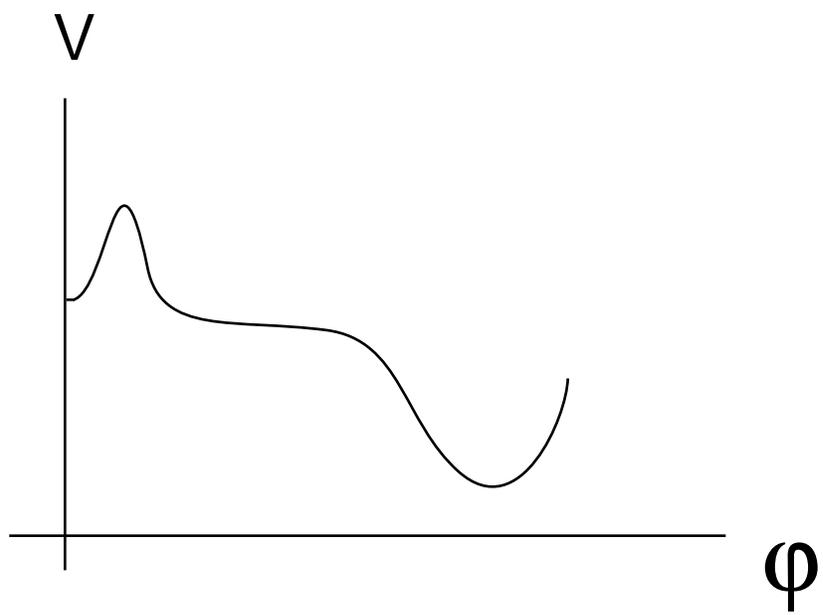
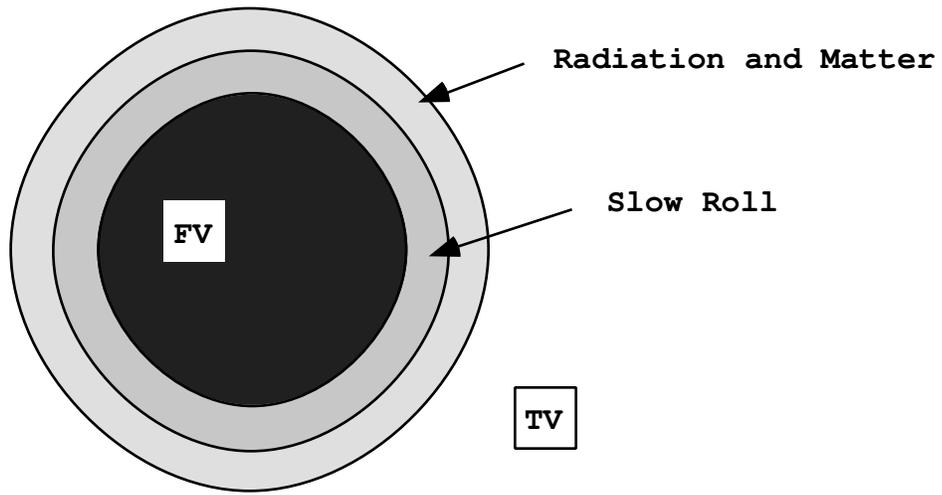
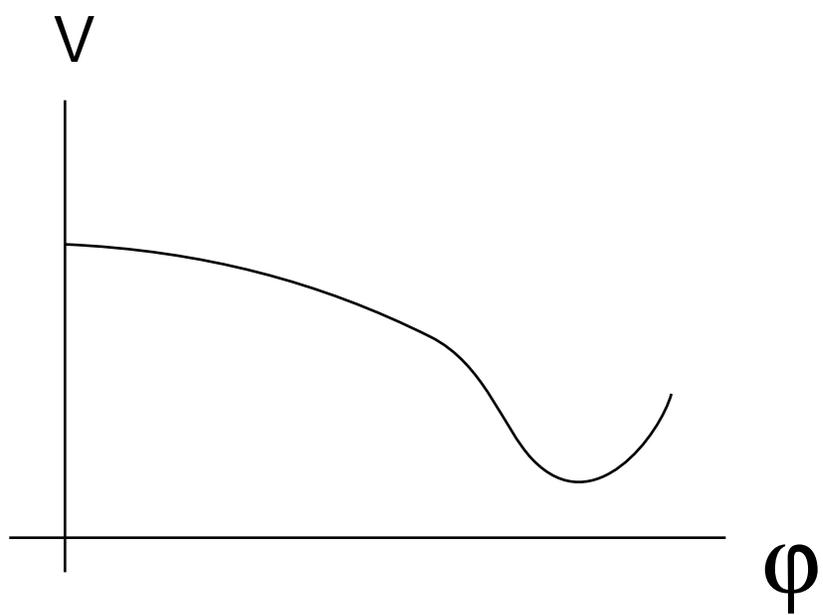
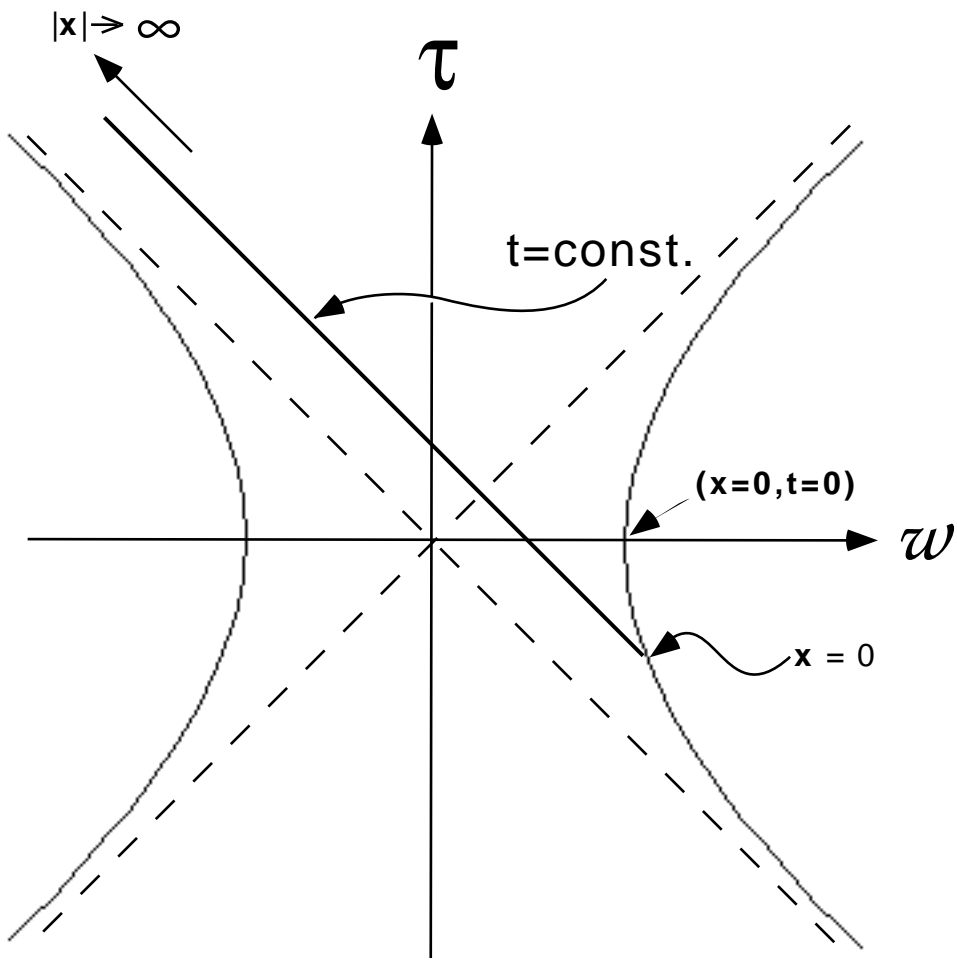
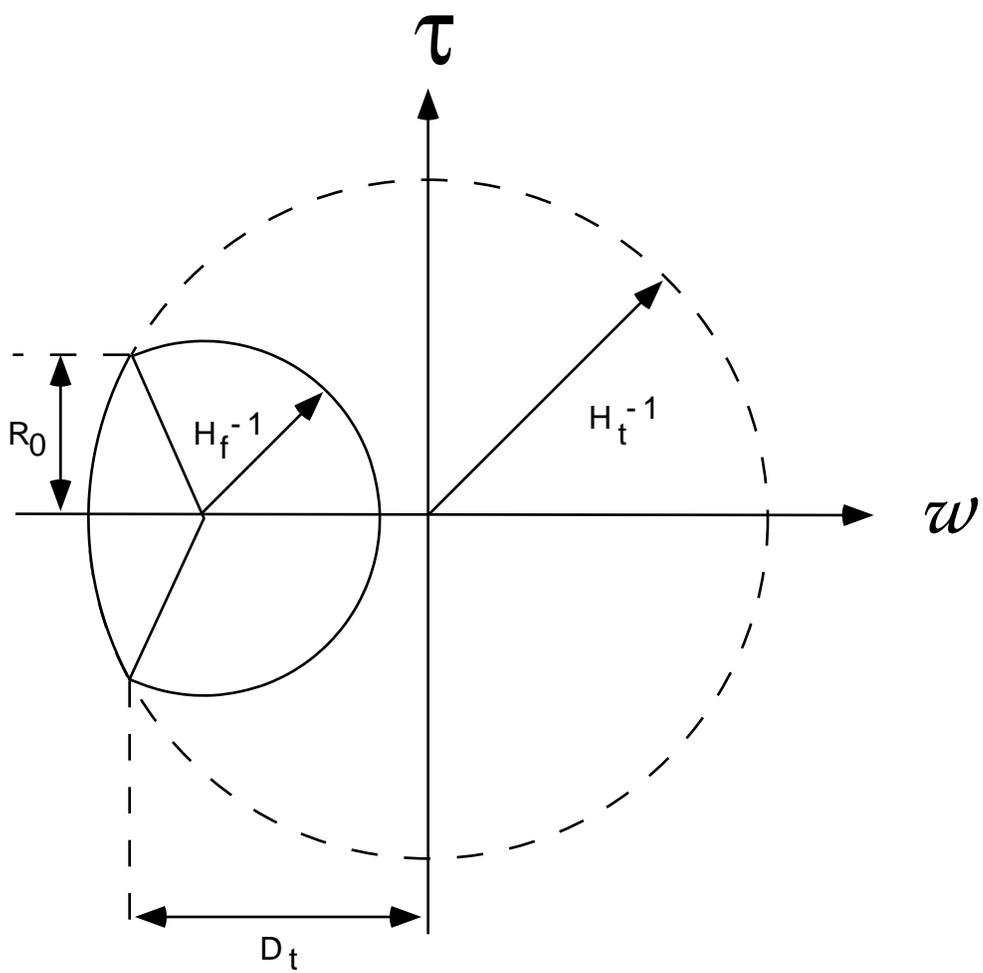
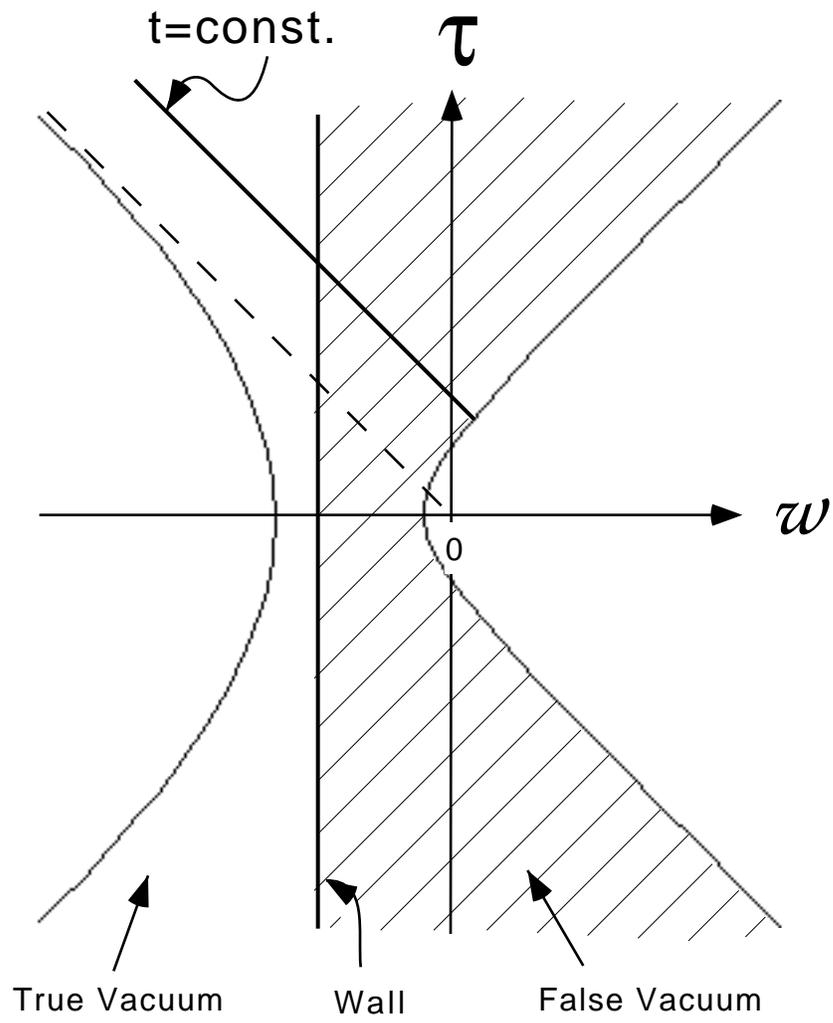
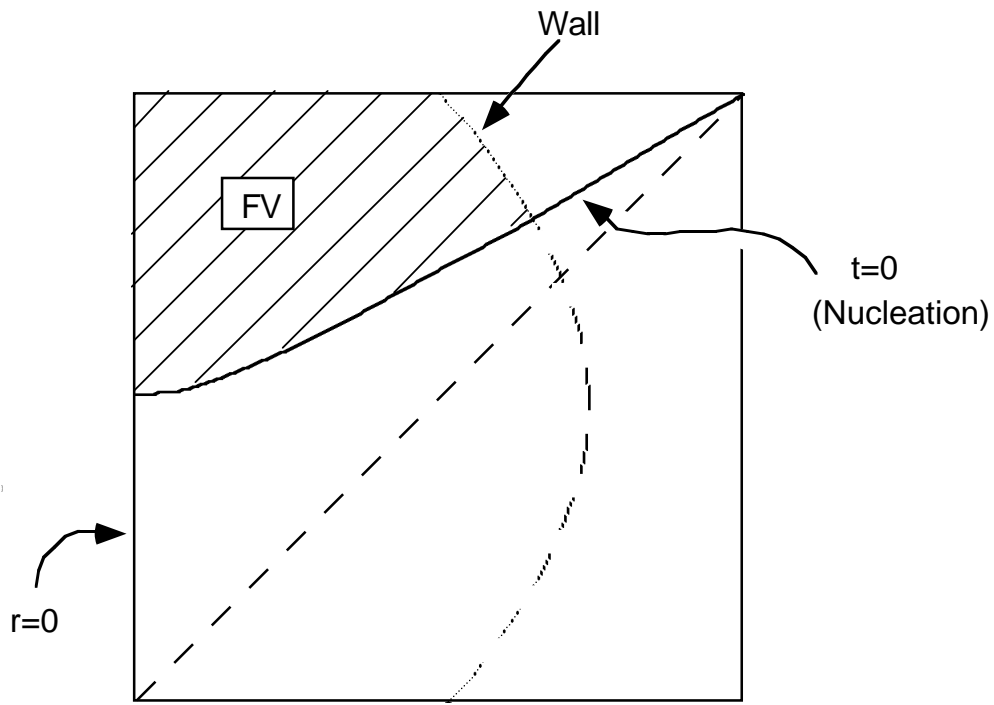
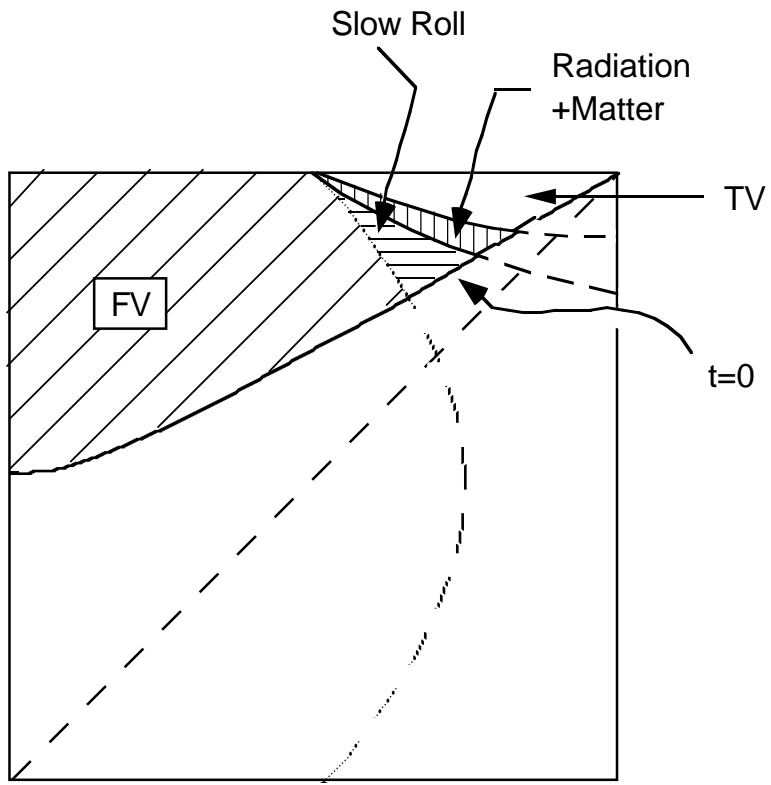
\end{document}